\documentclass[10pt,journal,compsoc]{IEEEtran}

\usepackage{amsmath,amsfonts}
\usepackage{array}
\usepackage{subfig}
\usepackage{textcomp}
\usepackage{stfloats}
\usepackage{url}
\usepackage{verbatim}
\usepackage{graphicx}

\newtheorem{definition}{\textbf{Definition}}
\usepackage{amssymb}
\usepackage[table,xcdraw]{xcolor}
\usepackage[ruled,vlined]{algorithm2e}

\ifCLASSOPTIONcompsoc
  \usepackage[nocompress]{cite}
\else
  \usepackage{cite}
\fi
\ifCLASSINFOpdf
\else
\fi
\begin{document}
%
\title{Don't Forget Too Much: Towards Machine Unlearning on Feature Level}
%
%
%
%

\author{Heng Xu,~
        Tianqing Zhu*,~\IEEEmembership{Member,~IEEE,} 
        Wanlei~Zhou,~\IEEEmembership{Senior Member,~IEEE,}
        and~Wei~Zhao,~\IEEEmembership{Fellow,~IEEE}
        \IEEEcompsocitemizethanks{
        \IEEEcompsocthanksitem *Tianqing Zhu is the corresponding author.
        \IEEEcompsocthanksitem Heng Xu and Tianqing Zhu are with the Centre for Cyber Security and Privacy and the School of Computer Science, University of Technology Sydney, Ultimo, NSW 2007, Australia (e-mail: heng.xu-2@student.uts.edu.au; tianqing.zhu@uts.edu.au).
        \IEEEcompsocthanksitem Wanlei Zhou is with the City University of Macau, Macau (e-mail: wlzhou@cityu.edu.mo). 
        \IEEEcompsocthanksitem Wei Zhao is with Shenzhen Institute of Advanced Technology, University of Chinese Academy of Sciences, Shenzhen, China (e-mail: zhao.wei@siat.ac.cn).
        }
        }

%
%

\markboth{Journal of \LaTeX\ Class Files,~Vol.~14, No.~8, August~2015}%
{Shell \MakeLowercase{\textit{et al.}}: Bare Demo of IEEEtran.cls for Computer Society Journals}
%



\IEEEtitleabstractindextext{%
\begin{abstract}
 Machine unlearning enables pre-trained models to remove the effect of certain portions of training data. Previous machine unlearning schemes have mainly focused on unlearning a cluster of instances or all instances belonging to a specific class. These types of unlearning might have a significant impact on the model utility; and they may be inadequate for situations where we only need to unlearn features within instances, rather than the whole instances. Due to the different granularity, current unlearning methods can hardly achieve feature-level unlearning. To address the challenges of utility and granularity, we propose a refined granularity unlearning scheme referred to as ``feature unlearning". We first explore two distinct scenarios based on whether the annotation information about the features is given: feature unlearning with known annotations and feature unlearning without annotations. Regarding unlearning with known annotations, we propose an adversarial learning approach to automatically remove effects about features. For unlearning without annotations, we initially enable the output of one model's layer to identify different pattern features using model interpretability techniques. We proceed to filter features from instances based on these outputs with identifying ability. So that we can remove the feature impact based on filtered instances and the fine-tuning process. The effectiveness of our proposed approach is demonstrated through experiments involving diverse models on various datasets in different scenarios.
\end{abstract}

}

\maketitle

\IEEEdisplaynontitleabstractindextext

%
\IEEEpeerreviewmaketitle

\IEEEraisesectionheading{\section{Introduction}\label{sec:introduce}}

%
%
%
%
\IEEEPARstart{M}{achine} unlearning refers to the process of eliminating the influence of specific training data on a machine learning model~\cite{DBLP:conf/sp/CaoY15,10.1145/3603620}. This technological advancement has gained significant attention recently, largely due to various factors, such as the \textit{the right to be forgotten} in regulations and laws,  privacy considerations, and the enhancement of model utility~\cite{webpage:GDPR,webpage:CCPA,DBLP:conf/sp/ShokriSSS17}. The conventional approach to machine unlearning involves retraining the model from scratch after deleting the unlearning instances. However, given the vast amount of training datasets used in modern machine learning, this method becomes impractical due to excessively high computational and time costs~\cite{DBLP:conf/sp/CaoY15}. To address this challenge, recent research has introduced a range of alternative techniques~\cite{DBLP:conf/sp/BourtouleCCJTZL21,DBLP:conf/ccs/Chen000H022,10.1145/3485447.3511997,DBLP:conf/icml/GuoGHM20}. Bourtoule et al.~\cite{DBLP:conf/sp/BourtouleCCJTZL21} proposed SISA (Sharded, Isolated, Sliced, and Aggregated), which based on the idea of data segmentation. Guo et al.~\cite{DBLP:conf/icml/GuoGHM20} calculated the effects of instances and adjusted model parameters to offset those effects.

Current machine unlearning mainly focuses on removing the effect at either the instance or class level. Instance-level unlearning approaches deal with the request to unlearn information related to an individual or a group of instances~\cite{DBLP:conf/sp/BourtouleCCJTZL21,DBLP:conf/ccs/Chen000H022,10.1145/3485447.3511997,DBLP:conf/icml/GuoGHM20}, while class-level unlearning methods involve eliminating the influences associated with all instances from a particular class~\cite{DBLP:conf/aaai/GravesNG21,DBLP:conf/ijcai/0001IIM21,DBLP:journals/ml/BaumhauerSZ22,DBLP:conf/www/Wang0XQ22}. However, many scenarios require unlearning in a feature level, in which the model owner only removes the impact of a particular feature, such as color, from the model. 
For instance, considering sensitive features such as age, gender, or skin color, which could be integrated across all instances in any model. The machine unlearning schemes at the instance level have to unlearn all instances containing those features, which is unpractical~\cite{DBLP:conf/ndss/WarneckePWR23}. Conversely, a feature level unlearning is urgently needed. 

\begin{figure*}[!t]
    \centering
    \subfloat{\includegraphics[width=1\textwidth]{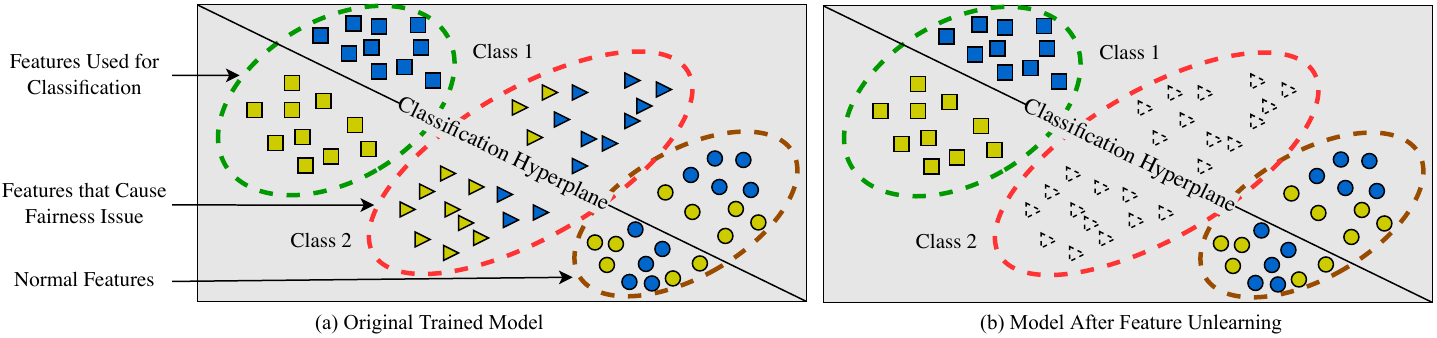}\label{fig:difference_a}}
    \caption{Feature unlearning helps to tackle the fairness problem. In each sub-figure, elements of different shapes represent different features; and different colors represent different feature annotations.  Figure 1a shows that the inconsistent frequency of features (triangles) may lead to unfairness. For example, instances possessing the yellow triangle feature will have a greater likelihood of being classified as class $2$, while instances possessing the blue triangle would be classified as $1$. Fairness solutions typically aim to mitigate bias by equalizing the frequency of biased features across different classes, with the goal of transforming these features into normal features. Since these features do not play vital roles for classification, we can eliminate the bias by directly unlearning those features and retaining the remaining features for the model classification~(as shown in Figure 1b).}
    \label{fig:difference}
\end{figure*}

Furthermore, the feature unlearning technique can serve as an alternative solution for alleviating model bias.  As shown in Figure~\ref{fig:difference}, the solutions to fairness issues primarily focus on equalizing the frequency of various features appearing across different classes in a dataset~\cite{Lim_2023_CVPR,DBLP:conf/iccv/WangZYCO19}. Feature unlearning can directly remove the features that lead to bias, without significant effect on model performance. 

However, compared with unlearning instance or class, unlearning feature, especially for unstructured data, is more challenging due to the complexity of instances. We have identified three distinct challenges if achieving feature unlearning. 
\begin{itemize}
    \item Firstly, a single instance may contain multiple tightly embedded features. The interactions between those multiple features and the model's processing across different layers are quite complex, making it difficult to eliminate the influence of a specific feature without negatively impacting other features. 
    
    \item Second, in many cases, we don't know the value of a feature, which we define as the \emph{annotation} of a feature. For example, an annotation of a skin color feature might be white. The unknown annotation means that the dataset does not provide any information about what these features look like. In a trained gender recognition model, the training dataset may only give information about the gender of each instance, without providing explicit information about their skin color. In this situation, unlearning skin color needs to remove the relevant features without any explicit feedback from the dataset, which is challenging. 
    
    \item Third, effectively evaluating the actual unlearning of features from the model is a significant challenge. Current instance-level evaluation metrics, such as accuracy-based or membership inference attack-based techniques ~\cite{DBLP:conf/nips/GuptaJNRSW21,DBLP:conf/mm/ZhangBHX22,DBLP:journals/tifs/ChundawatTMK23}, are unsuitable for assessing feature-level unlearning due to diverse granularity of unlearning targets. 
\end{itemize}

Current existing unlearning methods usually rely on data segmentation~\cite{10.1145/3485447.3511997,DBLP:conf/ccs/Chen000H022,DBLP:conf/sp/BourtouleCCJTZL21} or influence functions~\cite{DBLP:conf/cvpr/GolatkarAS20,DBLP:conf/icml/GuoGHM20} techniques, aiming to effectively partition the dataset to speed up model retraining or compute the effects of individual features, which cannot tackle above challenges. 

This paper introduces two feature unlearning schemes: feature unlearning with known annotations and feature unlearning without annotations. 
Based on this categorization, we propose two fine-tuning-based methodologies to address the above unlearning scenarios respectively. The fine-tuning scheme enables the model to quickly unlearn previously learned knowledge when undergoing fine-tuning for a new task~\cite{DBLP:conf/aaai/GravesNG21,DBLP:conf/ijcai/0001IIM21}. It is more efficient compared to the retraining-based unlearning strategies~\cite{DBLP:conf/sp/BourtouleCCJTZL21,DBLP:conf/ccs/Chen000H022,10.1145/3485447.3511997}. In our schemes, The known annotation uses adversarial training; and unlearning without annotations uses filtered instances to fine-tune model. 

To tackle the first challenge, in unlearning with known annotations, where we have annotations associated with the features to be unlearned, we employ adversarial training to fine-tune model. This technique automatically identifies and separates the feature information, allowing us to retain as many task-specific features as possible while removing the unlearning features in each instance. In the case of unlearning without annotations, we enable the output of one CNN model's layer to identify and separate distinct pattern features. This is achieved through model interpretability techniques. 



To address the second challenge in feature unlearning without annotations, we introduce a novel loss term~\cite{DBLP:conf/ijcai/0002WHZFZZ21} that encourages one CNN model's layer to identify distinct pattern features. To broaden the range of identification, we incorporate the eigengap heuristic technique, allowing for identifying a larger number of features while maintaining the desired effect. By identifying various feature information, we encode the instance to filter out the feature information that needs to be unlearned. Subsequently, we fine-tune the model to achieve the unlearning purpose based on those encoded instances. In contrast, when dealing with unlearning with known annotations, where the annotations are already known, we directly unlearn the features using adversarial learning.



To tackle the third challenge,
we evaluate the effectiveness of feature unlearning using the correlation between feature information and model task, as well as the variations of model accuracy. 
In addition, we qualitatively evaluate our two unlearning schemes by using Guided backpropagation~\cite{DBLP:journals/corr/SpringenbergDBR14} and visualizing whether our model has been fine-tuned on the revoked features for our two unlearning schemes.



In summary, we make the following contributions:

\begin{itemize}
    \item  We tackle the machine unlearning problem at the feature level and formally define two types of machine unlearning requests: feature unlearning with known annotations and feature unlearning without annotations.
    
    \item We introduce adversarial learning techniques to facilitate the feature unlearning with known annotations, while retaining useful feature information for the model task.
    
    \item We introduce model interpretability techniques to empower the output of one model's layer to decouple and identify various pattern features. These outputs, equipped with identifying capabilities, are then employed to facilitate unlearning without annotations. 
    \item We evaluate our feature unlearning from both quantitative and qualitative perspectives. In addition to evaluating based on accuracy, we also propose two new methods, including variation in accuracy and a gradient visualization-based scheme.

\end{itemize}

\section{Related Work}
\label{sec:related}

In response to \textit{the right to be forgotten}, the machine learning community has proposed several unlearning schemes. In our previously published survey paper~\cite{10.1145/3603620}, we conducted a comprehensive survey encompassing recent studies on machine unlearning techniques. This survey delved into several crucial aspects, including: (i) the motivation behind machine unlearning, (ii) the objectives and desired outcomes of unlearning, (iii) a novel taxonomy for categorizing existing machine unlearning techniques based on their rationale and strategies, and (iv) approaches for verifying the effectiveness of machine unlearning. In general, existing unlearning solutions mainly focus on the unlearning requests at the class or instance levels.

Class-level unlearning schemes refer to removing effects about all instances belonging to a specific class. Graves et al.~\cite{DBLP:conf/aaai/GravesNG21} presented a framework based on random relabeling and fine-tuning strategies for class-level unlearning. Takashi et al.~\cite{DBLP:conf/ijcai/0001IIM21} proposed an unlearning method, called selective forgetting, for lifelong learning environments. The authors devised mnemonic codes, which are synthetic signals tailored to specific classes that need to be unlearned. By leveraging the mechanism of catastrophic forgetting, these mnemonic codes were utilized to selectively unlearn particular classes without relying on the original data. Baumhauer et al.~\cite{DBLP:journals/ml/BaumhauerSZ22} first transformed the existing logit-based classifiers into the integrated model, consisting of a nonlinear feature extractor followed by logistic regression. Four filtration methods were then introduced to facilitate the unlearning of a specific class: naive unlearning, normalization, randomization, and zeroing. These methods filter the outputs of the logistic regression layer about the unlearning class. Wang et al.~\cite{DBLP:conf/www/Wang0XQ22} addressed the problem of class-level unlearning in federated learning environments based on Term Frequency-Inverse Document Frequency (TF-IDF) and fine-tuning techniques. However, the effectiveness of the above-mentioned schemes is limited in the scope of classes, consequently imposing constraints on their potential for broader scalability.

Instance-level unlearning schemes address the request to unlearn information pertaining to one instance or a cluster of instances from the model. Cao et al.~\cite{DBLP:conf/sp/CaoY15} transformed machine learning algorithms into summation representation and utilized these summations to develop statistical query learning models. To perform unlearning, affected summations, which contain instances requiring unlearning, were recalculated. Since only a few machine learning algorithms can be implemented as statistical query learning, this approach is unsuitable for complex models. Bourtoule et al.~\cite{DBLP:conf/sp/BourtouleCCJTZL21} presented a ``Sharded, Isolated, Sliced, and Aggregated” (SISA) architecture, similar to the existing ensemble training methodologies, to achieve the unlearning purpose. Other similar schemes are used in recommendation tasks~\cite{10.1145/3485447.3511997} and graph classification tasks~\cite{DBLP:conf/ccs/Chen000H022}. However, as the amount of unlearning data increases, those schemes will cause degradation in model performance, making them only suitable for small-scale scenarios~\cite{DBLP:conf/sp/BourtouleCCJTZL21}. In addition, smaller shards can decrease the retraining data size and reduce the costs, but it also impacts the performance of unlearned  model~\cite{10.1145/3485447.3511997,DBLP:conf/ccs/Chen000H022}. The unlearning methods adopted in \cite{DBLP:conf/icml/GuoGHM20,DBLP:conf/cvpr/GolatkarAS20,DBLP:conf/iwqos/LiuMYWL21} directly adjust the model parameters to offset the influence caused by instances, thereby achieving the unlearning purpose. The estimation of the influence caused by a particular training instance on the model parameters continues to be a challenging task when facing complex models. The existing theoretical schemes predominantly concentrate on simpler convex learning problems, such as linear or logistic regression. Schelter et al.~\cite{DBLP:conf/sigmod/SchelterGD21}, and Brophy et al.~\cite{DBLP:conf/icml/BrophyL21} considered the effective unlearning methods in tree-based models, while Nguyen et al.~\cite{DBLP:conf/nips/NguyenLJ20} focused on Bayesian model.

However, the above machine unlearning schemes primarily focused on instance-level requests. They are unable to handle more fine-grained unlearning requests at feature level. Alexander et al.~\cite{DBLP:conf/ndss/WarneckePWR23} provided a feature unlearning scheme based on influence functions, which allows feature unlearning through closed-form updates of model parameters~\cite{DBLP:conf/icml/GuoGHM20}. However, their scheme primarily addresses tabular datasets with enormous assumptions and does not consider more complex datasets like image data. In comparison, we aim to extend the unlearning capability to image data types, which poses significant challenges compared to current state-of-the-art schemes. Additionally, we also explore how to achieve unlearning purposes without relying on feature annotations for supervision.

\begin{table}
  \caption{Notations}
  \renewcommand{\arraystretch}{1.2}
  \label{tab:notations}
  \centering
  \begin{tabular}{c|l}
    \hline
    Notations &  Explanation \\
    \hline
    $\mathcal{D}$                       &The training dataset\\
    $\left( \mathbf{x}, y \right)$      &One instance in $\mathcal{D}$\\
    $M$                                 &The original trained model\\
    $\mathcal{D}_{u}$                   &Dataset that need to be unlearned\\
    $\mathcal{D}_{r}$                   &Remaining dataset\\
    $M_u$                               &Model after unlearning process\\
    $\mathrm{f}_i$                      &One feature in instance $\mathbf{x}$\\
    $A_i$                               &Different groups in target layer\\
    $s_{i,j}$                           &The similarity matrixes\\
    $\hat{\mathbf{x}}_i$                &The output of remover model\\
    $L_E$                               &The loss of adversarial learning\\
    $L(\mathbf{w}, \mathbf{A})$&The loss of identifier model\\
    $\beta,\lambda$                     &The hyperparameters in adversarial training\\
    $\gamma$                            &The hyper-parameters in identification training\\
    \hline
    \end{tabular}
\end{table}

\section{Preliminaries}


In the context of machine unlearning, we define the subset $\mathcal{D}_{u} \subset \mathcal{D}$ as a portion of the training dataset, whose influence we want to remove from the trained original model $M = \mathcal{A}(\mathcal{D})$, where $\mathcal{A}()$ is the training process. Conversely, the complement $\mathcal{D}_{r}=\mathcal{D}_{u}^{\complement} = \mathcal{D}/\mathcal{D}_{u}$ represents the dataset we intend to preserve those contributions within the model. Other important symbols that appear in this paper and their corresponding descriptions are listed in Table~\ref{tab:notations}.

With these definitions, we give the definition of machine unlearning for instance-level requests.

\begin{definition}[Machine Unlearning~\cite{DBLP:conf/sp/CaoY15}]
    \label{Definition:Machineunlearning}
    Let's define the unlearning process as $\mathcal{U}(M, \mathcal{D}, \mathcal{D}_{u})$, which aims to remove a cluster of instances $\mathcal{D}_{u}$ from the pre-trained model $M = \mathcal{A}(\mathcal{D})$. This unlearning process is a function that takes the pre-trained model $M$, the training dataset $\mathcal{D}$, and the unlearning dataset $\mathcal{D}_{u}$ as inputs and outputs a new model $M_u$. The objective of the unlearning process is to ensure that the unlearned model performs as if it had never seen the instances in the unlearning dataset.    
\end{definition}

If $\mathcal{D}_{u}$ in the above definition denotes all instances within a class, the above definition can be seen as class-level unlearning. However, class-level or instance-level unlearning schemes cannot handle feature-level unlearning requests. Feature unlearning focuses on the unlearning request on the feature level. Bau et al.~\cite{DBLP:conf/cvpr/BauZKO017} established a classification of six different types of semantics features within Convolutional Neural Networks (CNNs): objects, parts, scenes, textures, materials, and colors. Specifically, the first two categories can be broadly regarded as patterns related to objects characterized by specific shapes. On the other hand, the remaining four categories can be grouped as attribute patterns lacking distinct contours~\cite{DBLP:journals/pami/ZhangWWZZ21}. 

In this paper, we consider unlearning those two types of features, named \textit{pattern feature} and \textit{attribute feature}. For example, in the case of face recognition scenarios, the images typically contain various features, including but not limited to pattern features, such as the shape of the nose and the state of the mouth, and attribute features, such as male and youth. To represent the training dataset, we use $\mathcal{D}=$ $\left\{\left(\mathbf{x}_1, y_1, \mathbf{f}_1\right),\left(\mathbf{x}_2, y_2,\mathbf{f}_2\right), \ldots,\left(\mathbf{x}_n, y_n,\mathbf{f}_n\right)\right\} \subseteq \mathbb{R}^{d} \times \mathbb{R} \times \mathbb{R}^{k}$, where each $\mathbf{x}_{i} \in \mathcal{X}$ represents an individual instance, 
$y_{i} \in \mathcal{Y}$ is the corresponding label, and $n$ represents the size of the dataset $\mathcal{D}$. Unlike the general dataset definition, we define an extra item $\mathbf{f}_i$ in the dataset $\mathcal{D}$. $\mathbf{f}_i$ represents the feature annotations of this instance. Consider one instance $\mathbf{x}_i$ that consists of multiple features. The annotations of those features can be denoted as $\mathbf{f}_i = \left\{\mathrm{f}_1,\mathrm{f}_2, \ldots,\mathrm{f}_k \right\}$. For example, in the above face recognition scenarios, if the shape of the nose is described as whether it is pointy, $\mathrm{f}_{pointy\_nose}$ = True if it is, and $\mathrm{f}_{pointy\_nose}$ = False if it is not.

Now, we give the definition of feature unlearning. 

\begin{definition}[Feature Unlearning]
    \label{Definition:featureunlearning}
    Given one specific feature, the annotation is denoted as $\mathrm{f}_{i}$. We want to remove all effects of this feature from the trained model. Feature unlearning process $\mathcal{U}_{f}(M, \mathcal{D}, \mathrm{f}_{i})$ is defined as a function from an already-trained model $M = \mathcal{A}(\mathcal{D})$, a training dataset $\mathcal{D}$, and a specified feature $\mathrm{f}_{i}$ to a model $M_u$, which ensures that the unlearned model $M_{u}$ performs as though it had never seen the unlearning feature $\mathrm{f}_{i}$ in all training dataset.
\end{definition}

\section{Methodology}

\begin{figure*}[!t]    
    \centering
    \includegraphics[width=0.9\textwidth]{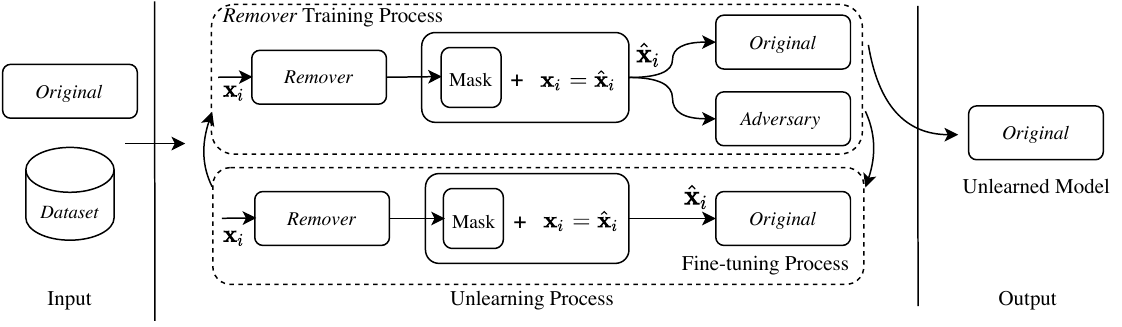}
    \caption{A schematic view of feature unlearning with known annotations.}
    \label{fig:schemeframework}
\end{figure*}

In this section, we first present an unlearning scheme based on adversarial training techniques. It is simple but effective in simultaneously unlearning multiple features with known annotations and maintaining the model's performance for the original task~(Section~\ref{subsection:featureunlearningiwthknownannations}). Moreover, we address the requests of unlearning features in scenarios where annotations are unknown and propose an unlearning method based on model interpretability and fine-tuning (Section~\ref{subsection:featureunlearningwithoutknownannotations} and Section~\ref{subsection:featureIdentificationwithoutknownannotations}). These approaches cater to both with known and without annotations cases, with adversarial learning for the former and interpretability-guided fine-tuning for the latter.

\subsection{Feature unlearning with known annotations}
\label{subsection:featureunlearningiwthknownannations}

In cases where the annotations of the features that need to be unlearned are known, we employ adversarial learning to fine-tune the original model and remove unlearning features automatically\footnote{In this paper, we use the term \textit{unlearning feature} to denote the feature that needs to be unlearned from the model.}. As shown in the middle in Figure~\ref{fig:schemeframework}, our scheme mainly consists of two sub-processes that run alternately with each other.  \textit{Remover training process} trains the \textit{remover}, whose purpose is to output a mask for filtering unlearning features and remaining useful features for original task. We classify features in each instance into two categories: \textit{target features} and \textit{task features}. Target features denote those features that are required to be unlearned, whereas task features encompass the ones linked to the task of the original model. As an example, in a scene recognition model, the features concerning the scene itself are considered task features, whereas the gender-related features that need to be unlearned are categorized as target features. We can further categorize target features according to whether it is pattern feature or attribute feature. The purpose of \textit{fine-tuning process} is to fine-tune the model based on the masked instances to achieve unlearning purpose.

\subsubsection{Remover training process} This process contains three main parts: \textit{remover}, \textit{original} and \textit{adversary} models. The upper right portion is the original model $M$, within which the feature unlearning operation needs to be performed, i.e., we need to remove the specified features from that model. Let $L\left(M\left(\mathbf{x}_i\right), y_i\right)$ denote the loss of the original model:
\begin{equation}
        \label{equation:1}
	\begin{split}
        &L_M = \sum L\left(M\left(\mathbf{x}_i\right), y_i\right)\\
        &~~~s.t. \left(\mathbf{x}_i, y_i\right) \in \mathcal{D}\\
	\end{split}
\end{equation}

The adversary $C$ is the lower right portion. It aims to predict whether the output instance $\hat{\mathbf{x}}_i$ from remover based on an input instance $\mathbf{x}_i$ contains target feature information. The adversary's objective is to minimize a loss that quantifies the amount of information about the target feature it can extract from $\hat{\mathbf{x}}_i$:
\begin{equation}
        \label{equation:2}
	\begin{split}
        &L_{C} = \sum L\left(C\left(\hat{\mathbf{x}}_i\right), \mathrm{f}_i\right)\\
        &~~~s.t. \left(\mathbf{x}_i, \mathrm{f}_i\right) \in \mathcal{D} \\
	\end{split}
\end{equation}

where the adversary, denoted as $C$, takes the outputs $\hat{\mathbf{x}}_i$ from the remover $E$ as its input, alongside the inputs $\mathbf{x}_i$ for remover. $\mathrm{f}_i$ represents the annotations for target feature of $\mathbf{x}_i$.

Remover $E$, encompasses an encoder and decoder framework designed to generate a mask. The purpose of this mask is to filter target features, while preserving task features as much as possible. In addition to this, to make the before and after filtered instances more similar, we introduce an extra loss term weighted by the parameter $\beta$ for remover:

\begin{equation}
    \label{equation:4}
	\begin{split}
        &L_E=\sum_i \left[ \beta\left|\mathbf{x}_i-\hat{\mathbf{x}}_i\right|_{\ell_1} + L\left(M\left(\hat{\mathbf{x}}_i\right), y_i\right)-\lambda L_{C}\right]\\
        &~~~s.t. \left(\mathbf{x}_i, \mathrm{f}_i, y_i\right) \in \mathcal{D}\\
	\end{split}
\end{equation}

The primary objective of the first term in Equation~\ref{equation:4} is to maintain a higher amount of information that already exists in the original instances. This helps to ensure that the instances after masking are as similar as possible to the original instance. The second term focuses on preserving a greater extent of task-specific feature information, which refers to task features. On the other hand, the third term eliminates the intended feature information, specifically the target features.

It is worth highlighting that we can address the scenario of unlearning multiple features simultaneously within this framework. That is, the number of feature $\mathrm{f}_i$ to be unlearned can be arbitrary. When only one feature is considered to be unlearned, the number of $\mathrm{f}_i$ can be one. For example, the gender feature information is unlearned in the scene recognition model. We can also unlearn multiple features at the same time. In such cases, we can use the following equation to evaluate the information related to multi-target features:

\begin{equation}
        \label{equation:3}
	\begin{split}
        &L_{C} = \sum L\left(C\left(\hat{\mathbf{x}}_i\right), (\mathrm{f}_1,\mathrm{f}_2,..,,\mathrm{f}_m)\right)\\
        &~~~s.t. \left(\mathbf{x}_i, (\mathrm{f}_1,\mathrm{f}_2,..,,\mathrm{f}_m)\right) \in \mathcal{D} \\
	\end{split}
\end{equation}

where $(\mathrm{f}_1,\mathrm{f}_2,..,,\mathrm{f}_m)$ represent multi features that need to be unlearned. $m$ is the number of those features.

\begin{algorithm}[!t]
	\small
	\caption{Unlearning with known annotations}
 	\label{algorithm:simple}
	\LinesNumbered 
	\KwIn{The full training set $\mathcal{D}$, original model $M$, remover $E$, adversary $C$, total iteration number $T$.}
	\KwOut{model $M_{u}$ after unlearning process}
  	~~~~~Training adversary $C$ based on the dataset $(\mathbf{x}_i, \mathrm{f}_{i}) \in \mathcal{D}$\\
        ~~~~~\For{$t=0; t<T; t++$}{
        ~~~~~\If{$i$ \% 2 == $0$}{
            ~~~~~Setting require\_grad = \textit{False} of all parameters from original model $M$ and adversary $C$.\\
            ~~~~~Setting require\_grad = \textit{True} of all parameters from remover $E$.\\
            ~~~~~\For{$(\mathbf{x}_i, \mathrm{f}_{i}, y_i)\in \mathcal{D}$}{
                ~~~~~Input $\mathbf{x}_i$ to remover $E$ and get $\hat{\mathbf{x}}_i$.\\
                ~~~~~$L_{C} = \sum L\left(C\left(\hat{\mathbf{x}}_i\right), \mathrm{f}_i\right)$.\\
                ~~~~~$L_M = \sum L\left(M\left(\hat{\mathbf{x}}_i\right), y_i\right)$.\\
                ~~~~~Updating remover $E$ based on $L_E=\sum_i \left[ \beta\left|\mathbf{x}_i-\hat{\mathbf{x}}_i\right|_{\ell_1} +  L_M-\lambda L_{C}\right]$.\\
                }
            }
        ~~~~~\Else{
            ~~~~~Setting require\_grad = \textit{False} of all parameters from remover $E$ and adversary $C$.\\
            ~~~~~Setting require\_grad = \textit{True} of all parameters from original model $M$.\\
            ~~~~~\For{$(\mathbf{x}_i, \mathrm{f}_{i}, y_i)\in \mathcal{D}$}{
                ~~~~~Input $\mathbf{x}_i$ to remover $E$ and get $\hat{\mathbf{x}}_i$.\\
                ~~~~~Updating original model $M$ based on $L_M = \sum L\left(M\left(\hat{\mathbf{x}}_i\right), y_i\right)$.\\
                }
            }
        }
	\Return {$M_{u} = M$}\\
\end{algorithm}

\subsubsection{Fine-tuning Process}

After one iteration of the remover training process is performed, we alternately fine-tune the model based on the filtered instance to gradually achieve the unlearning purpose:

\begin{equation}
        \label{equation:11}
	\begin{split}
        &L_M = \sum L\left(M\left(\hat{\mathbf{x}}_i\right), y_i\right)\\
	\end{split}
\end{equation}

After several iterations, target features will be unlearned from the original model. 
Algorithm~\ref{algorithm:simple} illustrates the whole unlearning process.

In line 1, we first train the adversary model to give it the ability to detect whether the specific feature information in an instance. Then, in lines 2-10, we train the remover $E$ to remove target features while retaining the relevant task features for the original model task. Specifically, in lines 4 to 5, we fix the parameters of the original model $M$ and adversary $C$ and only update remover $E$. In lines 6-10, we update the remover based on $L_E$. Alternately, as indicated in lines 12 to 16, we also fine-tune the original model based on instances generated by the remover. This effectively unlearns information about target features due to the presence of catastrophic forgetting~\cite{DBLP:conf/www/Wang0XQ22,DBLP:conf/aaai/GravesNG21}.

\subsection{Feature unlearning without annotations}
\label{subsection:featureunlearningwithoutknownannotations}

In the above-mentioned section, if we have annotations for the features that need to be unlearned, we can use adversarial training to achieve unlearning purpose. However, the adversarial training method cannot be applied to unlearn features if we lack annotations. This is because we cannot use the adversary model to remove features from instances.

Identifying and unlearning various features without annotations becomes extremely challenging. This task involves three main challenges. Firstly, how do we identify which features are present in the instance and can be unlearned? Secondly, how do we obtain more diverse features to meet different unlearning requirements? Thirdly, how do we unlearn feature information from the model when we know these features?  For the purpose of identifying various features, we modify a model so that its output of the middle layer can be used as an encoder to filter feature information. The modified model, we call it \textit{identifier model}. In this section, we illustrate how to achieve unlearning features without annotations based on the output of this identifier model. The construction of the identifier model will be introduced in the next Section.

\begin{figure*}[!t]
    \centering
    \includegraphics[width=0.95\textwidth]{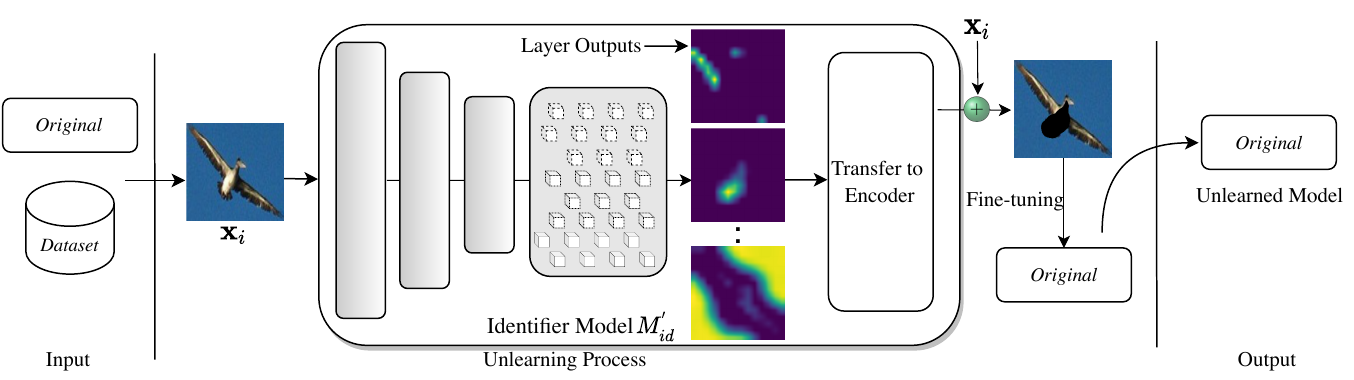}
    \caption{A schematic view of feature unlearning without annotations}
    \label{fig:Featureextraction}
\end{figure*}

Our unlearning scheme without annotations is shown in Figure~\ref{fig:Featureextraction}. We employ the output of trained identifier model as an encoder. The encoded instance is subsequently provided as input to the original model for fine-tuning. In the encoding process, we first feed one instance $\mathbf{x}_i$ into the trained identifier model and hook the output from a selected layer in \textit{identifier model}. The output of this layer can identify different pattern features, resulting in the output containing diverse feature information. It is worth noting that the identified features only contain different pattern features and no attribute features. Therefore, for feature unlearning without annotations, we only consider unlearning the pattern features. Following the above step, we choose a specific output based on the pattern features we intend to unlearn. This selected specific output will be used to encode instance $\mathbf{x}_i$. The algorithm outlining our approach is presented in Algorithm~\ref{algorithm:unlearning}.

In Algorithm~\ref{algorithm:unlearning}, we first fix the parameters of the identifier model and only update the original model~(Lines 1-2). Lines 3-9 involve obtaining the output of the identifier model, and subsequently encoding the original instance $\mathbf{x}_i$ based on these outputs. Specifically, when an output is chosen, it will be expanded to match the original instance's size~(line 6). Then, for each pixel point in the expanded output image, if its value is not zero, the corresponding pixel value in the original instance is removed to eliminate the feature information. These modified instances are then used to fine-tune the model further, resulting in feature unlearning without annotations~(line 11).


\begin{algorithm}[!t]
	\small
	\caption{Unlearning without annotations}
 	\label{algorithm:unlearning}
	\LinesNumbered 
	\KwIn{The full training set $\mathcal{D}$, original model $M$, model with identifying ability $M_{id}^{'}$, feature index $id$, total iteration number $T$.}
	\KwOut{model $M^{'}$ after unlearning process}
        ~~~~~Setting require\_grad = \textit{False} of all parameters from $M_{id}^{'}$.\\
        ~~~~~Setting require\_grad = \textit{True} of all parameters from $M$.\\
        ~~~~~\For{$(\mathbf{x}_i, y_i)\in \mathcal{D}$}{ 
            ~~~~~Input $\mathbf{x}_i$ to identifier model $M_{id}^{'}$. \\
            ~~~~~Hook output $O_i$ in target layer.\\
            ~~~~~Resize $O_i^{id}$ to size of $\mathbf{x}_i$.\\
            ~~~~~\For{each pixel $p\in O_i^{id}$ and $q \in \mathbf{x}_i$}{
            ~~~~~\If{$p$ is not equal to $0$}{
                    remove pixel $q$ in $\mathbf{x}_i$\\
                }
            }
        ~~~~~$\hat{\mathbf{x}}_i = \mathbf{x}_i$\\
        ~~~~~Finetuning $M$ based on $L_M = \sum L\left(M\left(\hat{\mathbf{x}}_i\right), y_i\right)$.\\
        }
    \Return {$M^{'} = M$}\\
\end{algorithm}

\subsection{Feature identification without annotations}
\label{subsection:featureIdentificationwithoutknownannotations}

\begin{figure}[!t]
      \centering
      \subfloat[Model Structure with selected layer]{\includegraphics[width=0.45\textwidth]{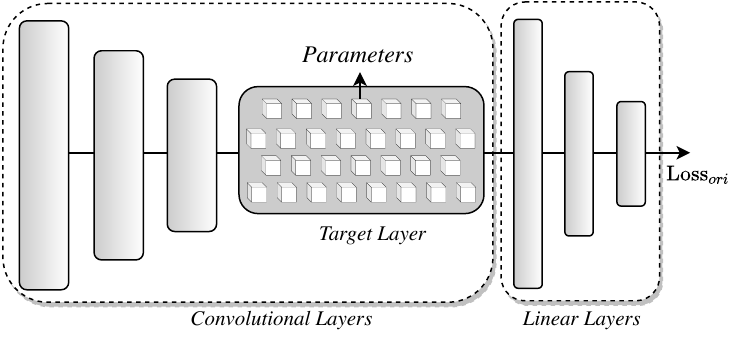}\label{fig:featureextraction_a}}\\
      \subfloat[Model Structure with grouped filter]{\includegraphics[width=0.45\textwidth]{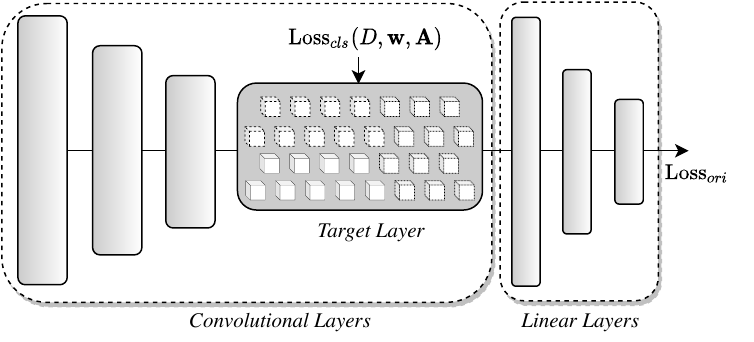}\label{fig:featureextraction_b}}
    \caption{Model structure for feature identification without annotations}
    \label{fig:featureextraction}
\end{figure}

To tackle the challenges of identifying which features can be unlearned, we empower the output of one model layer with the ability to identify different features. As illustrated in Figure~\ref{fig:featureextraction_a}, we opt for a specific layer within the model that can be strategically optimized to identify various feature information. Since filters in lower convolutional layers of CNNs typically capture basic texture features, while filters in higher convolutional layers are more inclined to encode object parts or complex pattern features. We focus on training models with automatic detecting features in higher convolutional layers~(gray layer in Figure~\ref{fig:featureextraction_a}), which we call \textit{target layer}. In addition, as shown in Figure~\ref{fig:featureextraction_b}, since CNNs usually use a set of filters to jointly detect specific image features rather than using a single filter~\cite{DBLP:conf/cvpr/FongV18}, we divide filters $\Omega={1,2, \cdots, d}$ in the target layer into distinct groups $A_1, A_2, \cdots, A_K$, where $A_1 \cup A_2 \cup \cdots \cup A_K=\Omega$; $A_i \cap A_j=\emptyset . \mathbf{A}=\{A_1, A_2, \cdots, A_K\}$ represents the filter partition. Based on this, we need to consider optimizing the selected target layer's parameters to identify various feature information.

Shen et al.~\cite{DBLP:conf/ijcai/0002WHZFZZ21} proposed a technique to optimize a convolutional layer's output to identify different feature information with the help of spectral clustering. The group number in the spectral clustering process represents the number of identified features. However, the optimization method in~\cite{DBLP:conf/ijcai/0002WHZFZZ21} needs to be given the expected number of groups $K$ in advance.  It does not consider how to obtain the maximum number of groups and how to ensure that the number of groups is optimal in the clustering process. We should obtain more diverse features in feature unlearning without annotations, enabling us to perform different feature unlearning operations.

To do that, we introduce the eigengap heuristic, which is based on perturbation theory and spectral graph theory, to determine the number of groups $K$~\cite{DBLP:journals/sac/Luxburg07,DBLP:conf/nips/Zelnik-ManorP04}. The eigengap heuristic indicates that the optimal value for $K$, representing the number of clusters in spectral clustering, is often identified by maximizing the difference between consecutive eigenvalues. A larger eigengap implies a stronger alignment between the eigenvectors of the desired outcome, resulting in improved performance of spectral clustering algorithms. Therefore, we use the loss in~\cite{DBLP:conf/ijcai/0002WHZFZZ21} to optimize our model and choose the value of $K$ based on the eigengap heuristic:

\begin{equation}
    \label{equation:5}
    L(\mathbf{w}, \mathbf{A})= L_{ori}(\mathcal{D}, \mathbf{w}) + \gamma L_{cls}(\mathcal{D}, \mathbf{w}, \mathbf{A})
\end{equation}

where $\mathbf{w}$ denotes the parameters of the model. $\mathbf{A}$ represents the parameters of each partition in the target layer. $L_{ori}(\mathcal{D}, \mathbf{w})$ denotes the original model loss on dataset $\mathcal{D}$:

\begin{equation}
    \label{equation:6}
    L_{ori}(\mathcal{D}, \mathbf{w})  = \frac{1}{n} \sum_{(\mathbf{x}_i, y_i) \in \mathcal{D}} L\left(\mathbf{x}_i, y_i, \mathbf{w}\right)
\end{equation}

$\gamma$ is a positive weight. $L_{cls}(\mathcal{D}, \mathbf{w}, \mathbf{A})$ is the new loss for identifying the feature information and can be defined as~\cite{DBLP:conf/ijcai/0002WHZFZZ21}:

\begin{equation}
    \label{equation:7}
	\begin{split}
        &L_{cls}(\mathcal{D}, \mathbf{w}, \mathbf{A})=-\sum_{k=1}^K \frac{S_k^{\text {within}}}{S_k^{\text {all }}}=-\sum_{k=1}^K \frac{\sum_{i, j \in A_k} s_{i j}}{\sum_{i \in A_k, j \in \Omega} s_{i j}}\\
        &~~~s.t.\\
        &~~~S_k^{\text {within}}=\sum_{i, j \in A_k} s_{i j}=\sum_{i, j \in A_k} \mathcal{K}\left(X_i, X_j\right) \\
        &~~~S_k^{\text {all}}=\sum_{i \in A_k, j \in \Omega} s_{i j}=\sum_{i \in A_k, j \in \Omega} \mathcal{K}\left(X_i, X_j\right)\\
        &~~~s_{i j}=\mathcal{K}\left(X_i, X_j\right)=\rho_{i j}+1 \geq 0\\
	\end{split}
\end{equation}

\begin{algorithm}[!t]
	\small
	\caption{Identifying different features}
 	\label{algorithm:identification}
	\LinesNumbered 
	\KwIn{The full training set $\mathcal{D}$, similarity metric $S_k^{\text {all}} = \{s_{i,j} | i,j \in A_k, k \in K\}$, target layer $l_{target}$, iteration number $T_{1}$ and $T_{2}$.}
	\KwOut{model $M_{id}^{'}$ after identifying process}
        ~~~~~Define and initiate one CNN model $M_{id}$\\
        ~~~~~eigenvalues, eigenvectors = $calculateeigen(S_k^{\text {all}})$\\
        ~~~~~$Top_{5} \leftarrow eigenDecomposition(eigenvalues)$ \\
        ~~~~~$K \leftarrow max(Top_{5})$ \\   
        ~~~~~Divide filters $\Omega$ in the target layer $l_{target}$ into distinct groups $A_1, A_2, \cdots, A_K$\\
        ~~~~~\For{$t=0; t<T_{1}; t++$}{
            ~~~~~$L_{ori}(\mathcal{D}, \mathbf{w})  = \frac{1}{n} \sum_{(\mathbf{x}_i, y_i) \in \mathcal{D}} \mathrm{L}\left(\mathbf{x}_i, y_i, \mathbf{w}\right)$\\
            ~~~~~Optimizing $M_{id}$ based on $L_{ori}(\mathcal{D}, \mathbf{w})$.\\

        }
        ~~~~~\For{$t=0; t<T_{2}; t++$}{
            ~~~~~$L_{cls}(\mathcal{D}, \mathbf{w}, \mathbf{A})=-\sum_{k=1}^K \frac{\sum_{i, j \in A_k} s_{i j}}{\sum_{i \in A_k, j \in \Omega} s_{i j}}$.\\
            ~~~~~$L_{ori}(\mathcal{D}, \mathbf{w})  = \frac{1}{n} \sum_{(\mathbf{x}_i, y_i) \in \mathcal{D}} \mathrm{L}\left(\mathbf{x}_i, y_i, \mathbf{w}\right)$\\
            ~~~~~$L(\mathbf{w}, \mathbf{A})= L_{ori}(\mathcal{D}, \mathbf{w}) + \gamma L_{cls}(\mathcal{D}, \mathbf{w}, \mathbf{A})$\\
            ~~~~~Optimizing $M_{id}$ based on $L(\mathbf{w}, \mathbf{A})$.\\

        }
    \Return {$M_{id}^{'} = M_{id}$}\\
\end{algorithm}

where $\rho_{i j}$ denotes the Pearson's correlation coefficient. $X_i$ denotes the set of output of the $i$-th filter. For calculating $K$, we initially calculate the Laplacian matrix using the provided similarity matrixes $S_k^{\text {all}}$. Subsequently, we calculate eigenvectors and eigenvalues from the Laplacian matrix. These eigenvalues are then used to calculate the consecutive difference and arranged in a sorted manner to identify the most optimal value of $K$. Our algorithm is shown in Algorithm~\ref{algorithm:identification}.

In line 2 in Algorithm~\ref{algorithm:identification}, we initially compute the eigenvalues and eigenvectors using the similarity values $S_k^{\text {all}}$. Subsequently, we determine the most suitable cluster numbers by identifying the indices associated with larger gaps between the eigenvalues. In this step, each selected cluster number $k$, satisfies the condition that all eigenvalues $\lambda_1, \ldots, \lambda_k$ are very small, but $\lambda_{k+1}$ is relatively large. Moving on to lines 4-5, we select the highest value as the number of clusters for filter division. This step ensures that we can achieve the maximum number of clusters while ensuring optimal clustering results. Finally, in lines 6-8, we first train the model using the $L_{ori}(\mathcal{D}, \mathbf{w})$ loss function. This initial phase ensures that the model achieves a basic level of classification performance. Subsequently, we proceed to train the identifier model further using the combined loss function~(lines 9-13). This step endows the model with the capacity to identify feature information effectively.

\section{Experiment}
In this section, we first evaluate our scheme from both qualitative and quantitative perspectives~(Section~\ref{section:Qualitative} and Section~\ref{section:quantitative}). The feasibility of the fine-tuning-based scheme and validity of introducing the eigengap heuristic technique is evaluated in Section~\ref{section:feasibilityanalysis} and Section~\ref{section:identifying}, respectively. We also test the effect of hyperparameters in Section~\ref{sec:effectparameters}. Finally, we consider one of the important applications of feature unlearning: as an optional alleviation scheme for model debiasing~(Section~\ref{section:featureunlearningapplication}). 

\subsubsection{Evaluation Metrics}
\label{section:metrics}
Training deep models with large datasets involves various randomness, which presents a challenge in evaluating the effectiveness of machine unlearning schemes. Existing instance-level evaluation methods primarily focus on model accuracy or attack-based techniques based on model performance ~\cite{DBLP:conf/nips/GuptaJNRSW21,DBLP:conf/mm/ZhangBHX22,DBLP:journals/tifs/ChundawatTMK23}. However, when it comes to unlearning specific features, the model itself cannot provide any performance information about those features, which results in the invalid of the above instance-level evaluation schemes. For example, precision information regarding male and female features in scene recognition classification models cannot be obtained. Therefore, these approaches are inadequate for evaluating whether feature information has been effectively unlearned from the model. To evaluate the effectiveness of feature unlearning, we consider three evaluation methods from different dimensions:

\begin{itemize}
    \item Accuracy from the adversary model: In our unlearning scheme with known annotations, the adversary has the ability to quantify the amount of information about the target feature. Therefore, we indirectly determine whether there is information about the features within the model based on the accuracy of the adversary.
    \item Variation in accuracy: For feature unlearning without annotations, it is difficult to quantitatively evaluate whether the pattern features are actually unlearned. We adopt a combination approach to determine whether features have been successfully unlearned, relying on the correlation between the target feature pattern and the task pattern, along with the accuracy of the model's task. Suppose we consider unlearning features $\mathrm{f}_i$ from any trained original model. The accuracy of the original model was $acc_{before}$ before the feature unlearning operation and became $acc_{after}$ after the unlearning operation. When the correlation between unlearning features and the original model task is low, the value of $acc_{after}-acc_{before}$ should be small. This indicates that when a feature that needs to be unlearned is unrelated to the model's task, unlearning the feature information does not significantly affect model accuracy. Conversely, when this feature is relevant to the model's classification task, unlearning that feature information will significantly decrease the model's accuracy. That is, the value of $acc_{after}-acc_{before}$ should be big.
    \item Gradient visualization: We introduce guided backpropagation~\cite{DBLP:journals/corr/SpringenbergDBR14} as our qualitative evaluation method. This approach primarily focuses on comprehending the features that impact the model's decisions or predictions. Employing this technique enables us to determine the presence or absence of the target feature visually.
\end{itemize}

\subsubsection{Baseline methods}
\label{section:baseline}
As we mentioned in Section~\ref{sec:related}, there is no relevant research on feature unlearning for image classification models. Feature unlearning for tabular data has been proposed in~\cite{DBLP:conf/ndss/WarneckePWR23}, however, this scheme cannot be extended to the image level and cannot be verified under a non-convex model that specific features in an image are really unlearned. Therefore, we consider the following methods as our baselines:
\begin{itemize}
   \item Instance-level fine-tuning: We remove all instances that contain target feature from the dataset ($\mathrm{f}_i$ = True) and use the remaining dataset to fine-tune the model for unlearning ($\mathrm{f}_i$ = False). This scheme is used to evaluate whether a model is able to unlearn feature information when it is fine-tuned with instances that do not contain the target features. If our adversarial learning scheme proves effective while the instance-level fine-tuning scheme does not, this further illustrates the validity of our approach. 
   \item Instance-level retraining: The traditional method for machine unlearning typically requires retraining the model from scratch after removing the instances that need to be unlearned. There are also numerous approaches proposed to enhance the retraining process, such as~\cite{DBLP:conf/sp/CaoY15,DBLP:conf/sp/BourtouleCCJTZL21,DBLP:conf/ccs/Chen000H022,10.1145/3485447.3511997}. To illustrate that traditional instance-level methods cannot achieve feature-level unlearning, we conducted this experiment as a comparative case.
\end{itemize}

In the scenario of unlearning without annotations, we identify which pattern features within instances can be unlearned by introducing the model interpretability technique. To identify more feature information and achieve optimal training results, we introduce the eigengap heuristic to optimize the identification process. As a comparison, we evaluate our scheme with icCNN proposed in~\cite{DBLP:conf/ijcai/0002WHZFZZ21}.

\subsection{Performance analysis: visualization results}
\label{section:Qualitative}

To evaluate the effectiveness of our scheme, we conduct experiments based on the metric outlined in Section~\ref{section:metrics}. We consider three different settings:

\begin{enumerate}
    \item Single feature unlearning with annotations: This scenario refers to the fundamental process of feature unlearning, wherein we aim to unlearn information associated with a particular feature from the model. During this unlearning request, we can access corresponding annotation information about the unlearning feature from the dataset. 
    \item Multi feature unlearning with annotations: As mentioned in Section~\ref{subsection:featureunlearningiwthknownannations}, our scheme can unlearn multiple feature information simultaneously. This scalability allows us to unlearn all features at once efficiently. To validate this capability, we conducted experiments involving the unlearning of multiple features concurrently.
    \item Feature unlearning without annotations: This aspect presents one of the most challenging requirements in feature unlearning since we lack knowledge about the annotations of the features that need to be unlearned. Without explicit information regarding the annotations, it becomes significantly more complex to identify and unlearn the features' information effectively.
\end{enumerate}

\begin{figure*}[!t]
    \centering
    \subfloat[Results for single feature]{\includegraphics[width=0.31\textwidth]{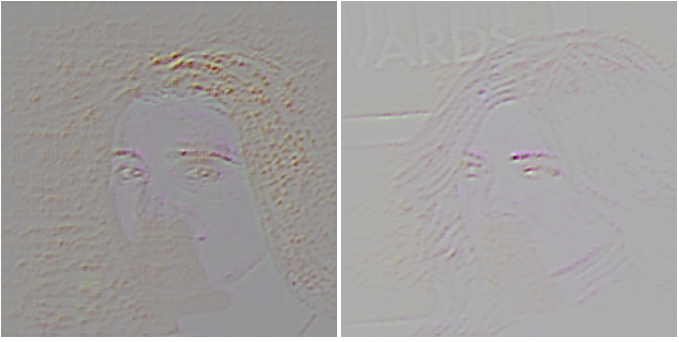}\label{fig:qualitative_a}}
    \subfloat[Results for multi feature]{\includegraphics[width=0.31\textwidth]{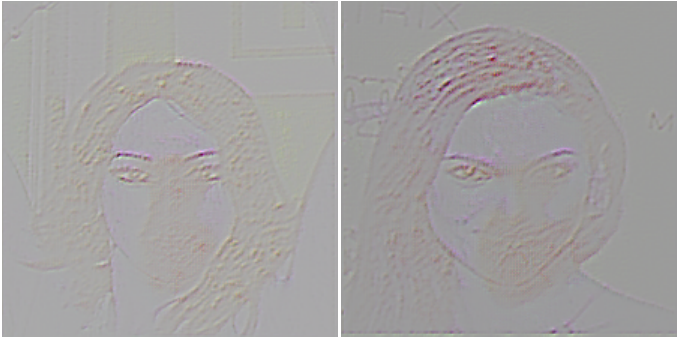}\label{fig:qualitative_b}}
    \subfloat[Results without annotations]{\includegraphics[width=0.31\textwidth]{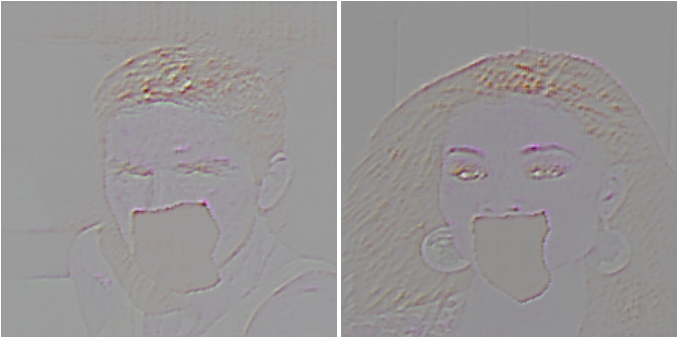}\label{fig:qualitative_c}}
    \caption{Qualitative results based on the guided backpropagation~\cite{DBLP:journals/corr/SpringenbergDBR14}. In all Figures, the gradient map lacks information about unlearning features. Figure~\ref{fig:qualitative_c} demonstrates the results without annotations, which illustrate that our interpretability-based unlearning scheme achieves almost the same results as the known annotations scheme (Figure~\ref{fig:qualitative_a}). All results show that both types of unlearning schemes can remove the gradient information about unlearning features during the unlearning process.
    }%
    \label{fig:qualitative}
\end{figure*}

\textbf{Setup.} For setting 1), we choose the ResNet architecture for both the original and adversary models and select U-net architecture for the remover model. In the first step, we respectively train the original model and the adversary model to identify \textit{Bald} and \textit{Mouth Slightly Open} tasks in CelebA. We aim to unlearn the information related to whether the mouth is open, specifically the \textit{Mouth Slightly Open} feature from the original model. For the training process, we set epoch = 10, batch size = 50 and learning rate = 0.000005. For executing the adversarial unlearning process, we set the epoch = 50, batch size = 36, learning rate = 0.000005, $\beta = 5.0$ and $\lambda = 5.0$. 

For setting 2), we choose the model structure from setting 1) for this setting. In order to simulate unlearning multiple features, we group the features \textit{Mouth Slightly Open} and \textit{Pointy Nose} together and aim to unlearn them from the original model. Concurrently, we train the original model to identify \textit{Bald} tasks in CelebA. The adversary model is trained to possess the capability of multi-task classification, allowing it to recognize both \textit{Mouth Slightly Open} and \textit{Pointy Nose} in CelebA. For training adversary model, we set epoch=20, batch size =128 and the learning rate=0.000005. For the original model, we set epoch = 10, batch size = 50 and learning rate = 0.000005. For executing the adversarial unlearning process, we set the epoch=50, batch size=36, learning rate=0.0001, $\beta = 1.0$ and $\lambda=10.0$.

For setting 3), we also use the ResNet model as the architecture of our identifier model and select the CelebA dataset to train this model with the ability to recognize different features. Afterward, we fine-tune the original model using the new encoded images to achieve the purpose of unlearning features without annotations. Specifically, we select the filter that recognizes the mouth in the identifier model and implement the unlearning \textit{Mouth} pattern feature in the \textit{Bald} and \textit{Smiling} classification models based on this filter. For training the identifier model, we first set epoch = 200, batch size = 256 and learning rate = 0.000001 to train it based on the loss $L_{ori}(\mathcal{D}, \mathbf{w})$. By following this, the model will attain a fundamental level of classification performance. After that, we continue to train the identifier model based on the loss $L_{cls}(\mathcal{D}, \mathbf{w}, \mathbf{A}) + L_{ori}(\mathcal{D}, \mathbf{w})$ and set the batch epoch = 2500, size = 128 and learning rate = 0.00001. This step endows the model with the capacity to identify feature information effectively. Other parameters in the training process are the same in~\cite{DBLP:conf/ijcai/0002WHZFZZ21}. For the fine-tuning unlearning process, we set the learning rate as 0.001, batch size = 128.

During the unlearning process of all the above experimental settings, we show the gradient based on the gradient visualization technique within the original model. The results are shown in Figure~\ref{fig:qualitative}.

\textbf{Results.} In Figure~\ref{fig:qualitative}, Figure~\ref{fig:qualitative_a} shows the results of unlearning a single feature with known feature annotations, while Figure~\ref{fig:qualitative_b} illustrates the result of unlearning multiple features. Figure~\ref{fig:qualitative_c} shows the result of unlearning feature without annotations. As can be seen from all Figures, the gradient map produced by guided backpropagation does not have information about features that need to be unlearned, which suggests that our scheme does not use feature-related information to fine-tune the model. Examples include the mouth information in Figures~\ref{fig:qualitative_a} and~\ref{fig:qualitative_c}, and the mouth and nose information in Figure~\ref{fig:qualitative_b}. Simultaneously, images unrelated to the unlearning features remain unaffected, indicating that adversarial training or encoding processes can preserve information unrelated to those unlearning features. Considering the impact of catastrophic forgetting, the model will autonomously unlearn information about specific features. In addition to this, it should be emphasized that the results in Figure~\ref{fig:qualitative_c} demonstrate the results of feature unlearning without annotations. It can be seen that even without annotations, our unlearning scheme based on interpretability techniques can achieve almost the same results as the known annotations scheme~(Figure~\ref{fig:qualitative_a}). Both unlearning schemes can remove the gradient information about the mouth feature.

\subsection{Performance analysis: quantitative results}
\label{section:quantitative}
In this section, we carry out the experiments utilizing the experimental setup detailed in Section~\ref{section:Qualitative}. In addition, as a comparison, we also try to remove feature information from the model based on \textit{instance-level fine-tuning} and \textit{instance-level retraining} schemes mentioned in Section~\ref{section:baseline}:

\begin{figure*}[!t]
      \centering
      \subfloat[Single Feature Unlearning]{\includegraphics[width=0.33\textwidth]{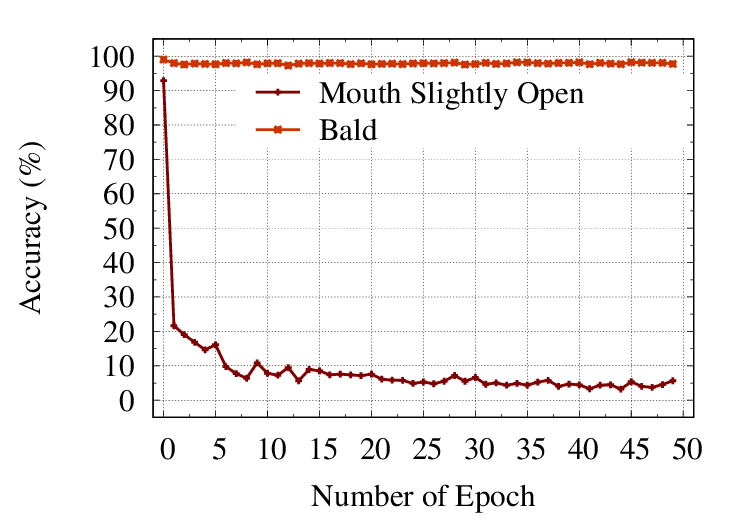}\label{fig:unlearningresults_a}}
      \subfloat[Multi Feature Unlearning]{\includegraphics[width=0.33\textwidth]{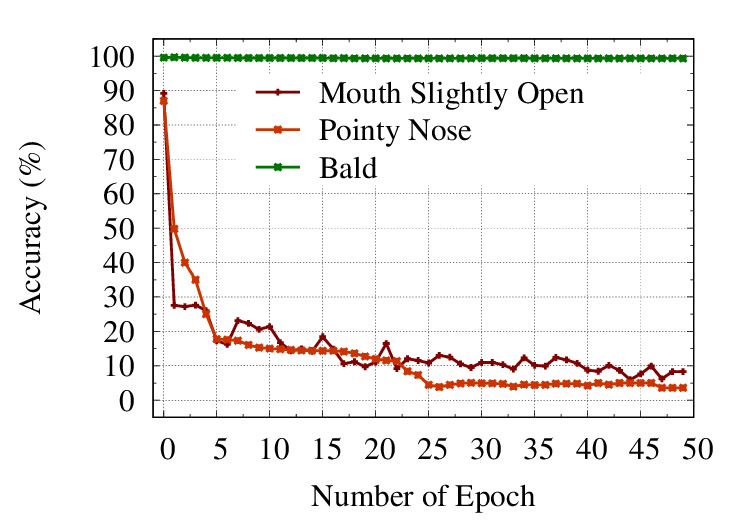}\label{fig:unlearningresults_b}}
      \subfloat[Feature Unlearning without Annotations]{\includegraphics[width=0.33\textwidth]{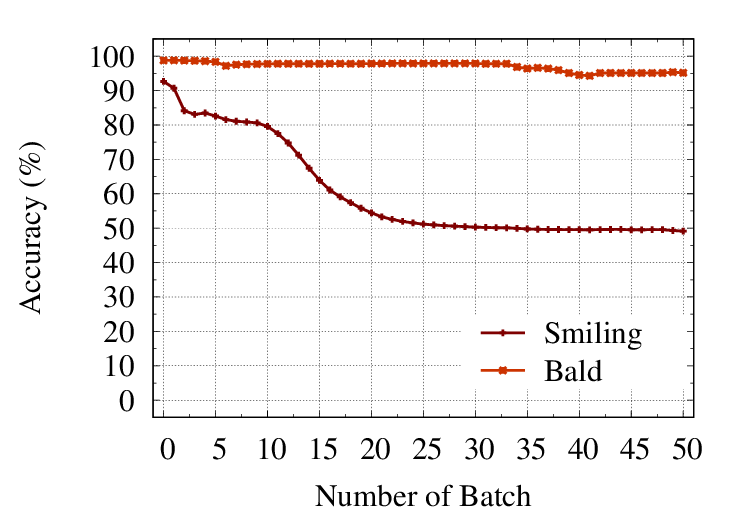}\label{fig:unlearningresults_c}}\\
      \subfloat[Instance-level finetuning (Single)]{\includegraphics[width=0.33\textwidth]{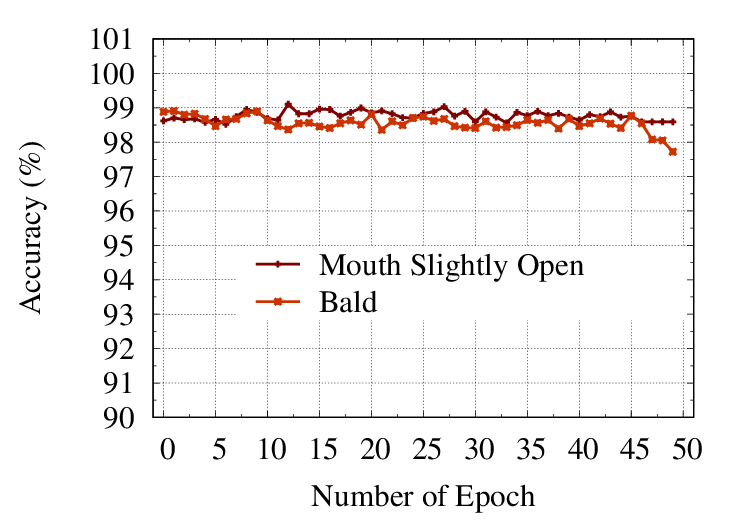}\label{fig:unlearningresults_d}}
      \subfloat[Instance-level finetuning (Multi)]{\includegraphics[width=0.33\textwidth]{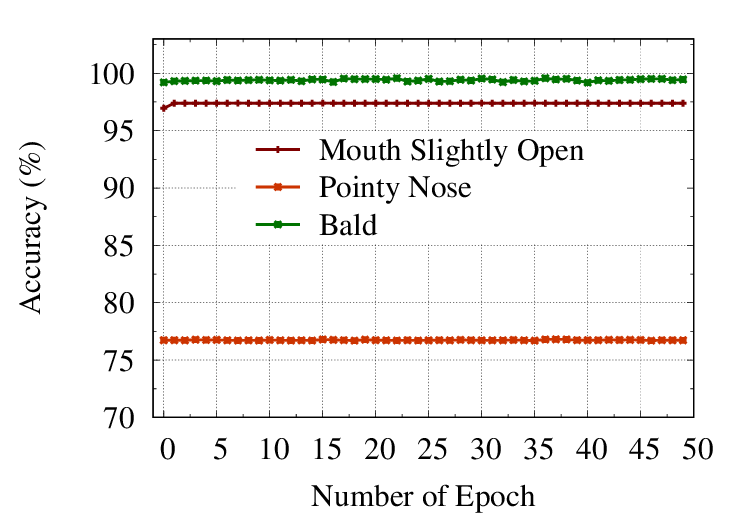}\label{fig:unlearningresults_e}}
      \subfloat[Instance-level retraining]{\includegraphics[width=0.33\textwidth]{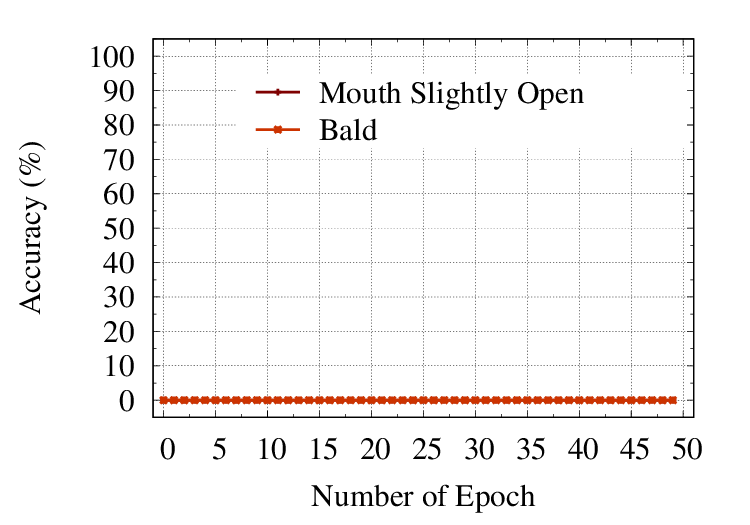}\label{fig:unlearningresults_f}}\\
    \caption{Quantitative results with different configurations. Precision for unlearning features in Figures~\ref{fig:unlearningresults_a} and~\ref{fig:unlearningresults_b} progressively declines during the unlearning process. Meanwhile, the model's original task performance remains minimally altered. For the results in~\ref{fig:unlearningresults_c}, since the features to be unlearned are not highly correlated with the bald, the accuracy of the bald is essentially unchanged. While the correlation with smiling is high, the accuracy of smiling changes significantly. Other solutions cannot achieve the unlearning purposes. All results highlight feature unlearning can retain task-related information while removing targeted features
    }%
    \label{fig:unlearningresults}
\end{figure*}

\begin{itemize}
    \item For instance-level fine-tuning, we choose the ResNet architecture for the original model and train the original model to identify \textit{Bald} tasks in CelebA. Then, we aim to unlearn the information related to whether the mouth is open, specifically the \textit{Mouth Slightly Open} feature from the original model. We delete all instances that \textit{Mouth Slightly Open} = True from the dataset and use the remaining dataset~(\textit{Mouth Slightly Open} = False) to fine-tune the original model for unlearning. In addition, we also consider unlearning \textit{Mouth Slightly Open} and \textit{Pointy Nose} features from the original model. For the training process of the original model, we set epoch = 10, batch size = 50 and lr = 5e-06. For the fine-tuning process, we set epoch = 50, batch size = 50 and lr = 1e-04.
    \item For instance-level retraining, we also select the ResNet architecture as the original model and retrain it to recognize \textit{Bald} tasks within CelebA. Our goal is to remove the information associated with the state of the mouth, particularly the ``Mouth Slightly Open" feature, from the original model. Therefore, we first remove all instances from the model that illustrate the state of whether the model is opening (Mouth Slightly Open = True and false), and then retrain the model to recognize \textit{Bald} tasks. For the retraining process, we set epoch=50, batch size = 50 and learning rate = 1e-04.
\end{itemize}

The results are then evaluated using the metrics mentioned in~\ref{section:metrics} and shown in Figure~\ref{fig:unlearningresults}.

\textbf{Results.} As illustrated in Figure~\ref{fig:unlearningresults}, Figure~\ref{fig:unlearningresults_a} and Figure~\ref{fig:unlearningresults_b} show the results of feature unlearning with known annotations, while Figure~\ref{fig:unlearningresults_c} illustrates the result of feature unlearning without annotations. The precision associated with the features that need to be unlearned in Figure~\ref{fig:unlearningresults_a} and Figure~\ref{fig:unlearningresults_b} exhibits a progressive decrease as the unlearning process goes on. Furthermore, the model's original task performance demonstrates marginal alteration, emphasizing the efficacy of feature unlearning in retaining task-related information while removing target feature information from the model. In Figure~\ref{fig:unlearningresults_c}, the accuracy concerning the feature bald remains relatively stable, whereas the accuracy about the feature smiling experiences a substantial reduction. This result is consistent with our expected findings, as the feature smiling exhibits substantial relevance to the features that need to be unlearned, whereas its correlation with the feature bald is less. As a result, the effect on the performance of smiling is more significant. The results demonstrate that our interpretability-driven unlearning approach can achieve nearly identical outcomes as the established annotation-based scheme, even in the absence of annotations. 

Figure~\ref{fig:unlearningresults_d} and Figure~\ref{fig:unlearningresults_e} show the results of instance-level fine-tuning with known annotations, while Figure~\ref{fig:unlearningresults_f} illustrates the result of instance-level retraining with known annotations. Contrary to our results in Figure~\ref{fig:unlearningresults_a} and Figure~\ref{fig:unlearningresults_b}, in Figure~\ref{fig:unlearningresults_d} and Figure~\ref{fig:unlearningresults_e}, the accuracy of the features to be unlearned does not decrease. This suggests that even when all instances containing \textit{Mouth Slightly Open} feature are removed~(\textit{Mouth Slightly Open} = True), the model can still obtain information from other mouth-related instances~(There's a mouth, but it's not opening.). Certainly, it's technically possible to remove all mouth-related instances from the dataset, but this would lead the model's original accuracy to zero, as illustrated in Figure~\ref{fig:unlearningresults_f}. For the experimental result in Figure~\ref{fig:unlearningresults_f}, since it removes all instances where the mouth is opening or closing~(both \textit{Mouth Slightly Open} = True and \textit{Mouth Slightly Open} = False), it results in no instances being available for model training during the retraining process, resulting in the accuracy of the model remaining at 0. 

From all the above results, our unlearning schemes can effectively eliminate the information associated with the target features from the model, while all instance-level either fail to remove completely or remove too much. In conclusion, our feature unlearning strategy effectively removes feature information from the models.

\begin{figure*}[!t]
      \centering
      \subfloat[Results of Original Training Process]{\includegraphics[width=0.25\textwidth]{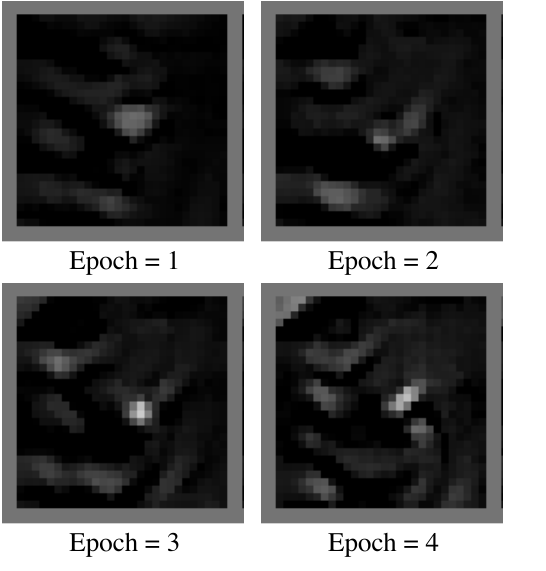}\label{fig:modelinversionattack_original}}
      \subfloat[Results of Retraining Process]{\includegraphics[width=0.268\textwidth]{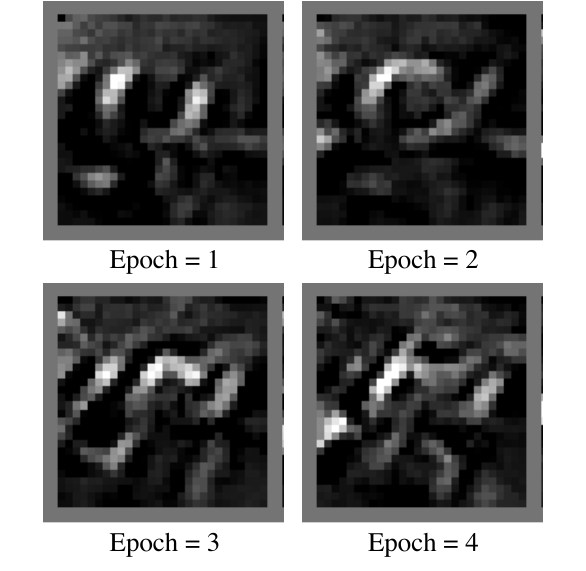}\label{fig:modelinversionattack_retraining}}
      \subfloat[Results of Finetuning Process]{\includegraphics[width=0.253\textwidth]{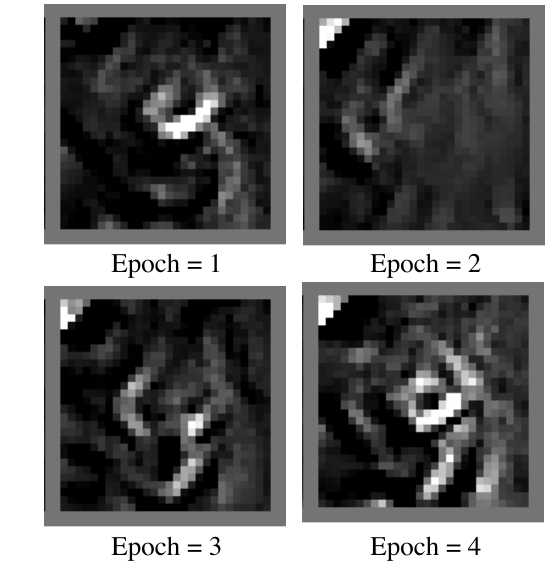}\label{fig:modelinversionattack_finetuning}}
    \caption{The results of model inversion attack. In Figure~\ref{fig:modelinversionattack_retraining}, the attack is almost ineffective due to the model lacking knowledge about the unlearning class during the retraining process. In Figure~\ref{fig:modelinversionattack_finetuning}, after the first round of the fine-tuning process, recovering unlearning class information gets harder due to limited information to rely on. As fine-tuning continues, information retention decreases, making recovery more difficult. All results show that fine-tuning scheme can also unlearn the information about the unlearning class.
    }%
    \label{fig:modelinversionattack}
\end{figure*}

\subsection{Comparing with learning from scratch}
\label{section:feasibilityanalysis}

In this section, we aim to verify whether the fine-tuning unlearning scheme yields a nearly equivalent unlearning outcome while exhibiting more efficiency in comparison to the completely retraining scheme~\cite{DBLP:conf/aaai/GravesNG21}. To accomplish this, we conduct separate evaluations to assess the effectiveness of the fine-tuning and retraining schemes. We employ the currently dominant validation schemes, model inversion attacks and membership inference attacks~(MIAs), to assess the results.

\begin{figure*}[!t]
      \centering
      \subfloat[MIAs in MNIST Dataset]{\includegraphics[width=0.33\textwidth]{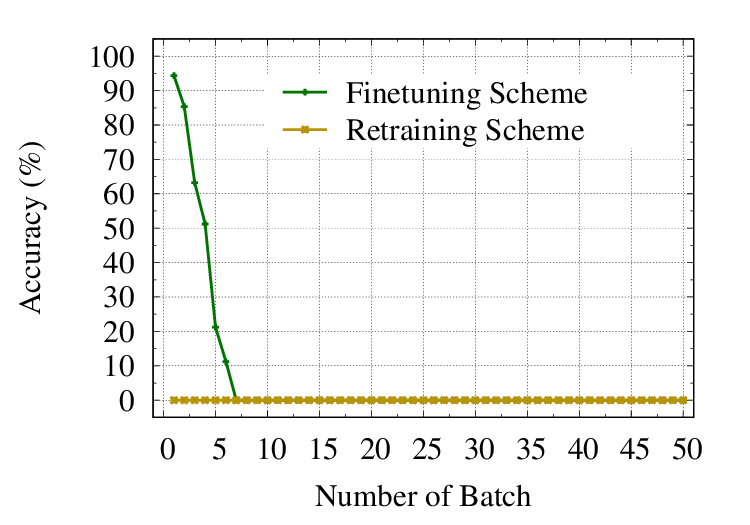}\label{fig:membershipinferenceattack_a}}\hfill
      \subfloat[MIAs in CIFAR-10 Dataset]{\includegraphics[width=0.33\textwidth]{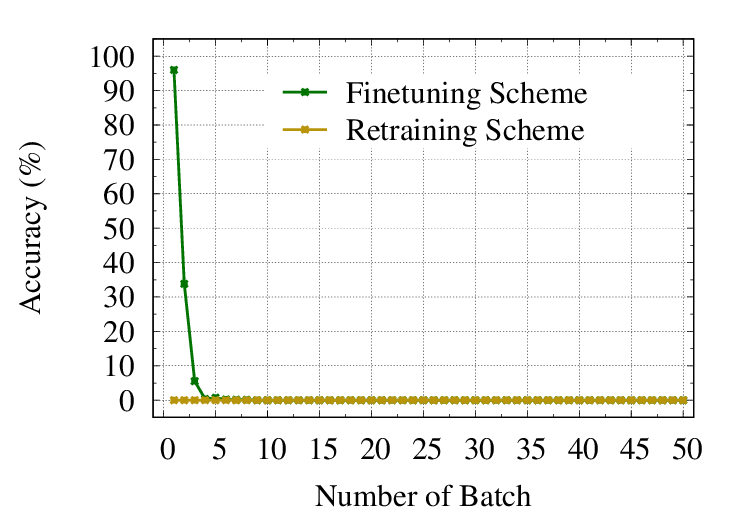}\label{fig:membershipinferenceattack_b}}\hfill
      \subfloat[Time Consume Evaluation]{\includegraphics[width=0.33\textwidth]{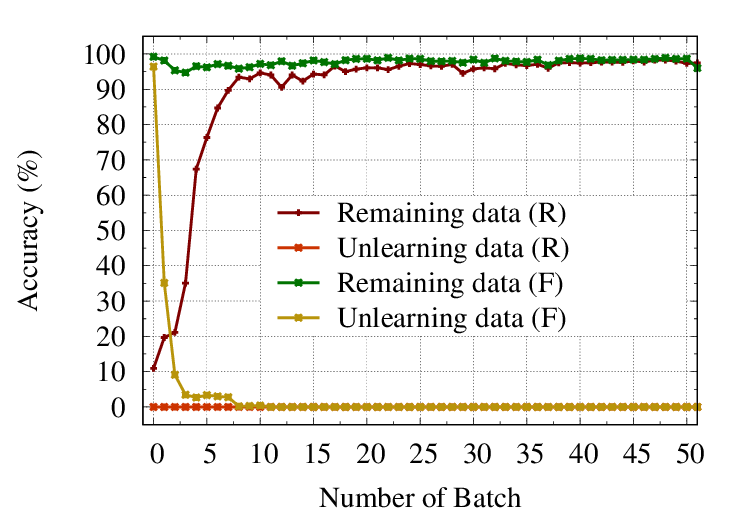}\label{fig:membershipinferenceattack_c}}
    \caption{The results of membership inference attacks. In Figure~\ref{fig:membershipinferenceattack_c}, \textit{R} is an abbreviation for retraining from scratch scheme, and \textit{F} means fine-tuning scheme. In Figure~\ref{fig:membershipinferenceattack_a} and~\ref{fig:membershipinferenceattack_b}, for all retraining schemes, MIAs have zero success rate, implying that it can't successfully derive the training dataset containing the unlearning class. For fine-tuning schemes, as the training progresses, the effectiveness of MIAs falls, indicating the fine-tuning scheme can also unlearn information about the unlearning class. In addition, in Figure~\ref{fig:membershipinferenceattack_c}, the fine-tuning scheme achieves much faster unlearning results compared to retraining the model to a usable state.
    }
    \label{fig:membershipinferenceattack}
\end{figure*}

\textbf{Setup.} We implement the model inversion attacks as described in~\cite{DBLP:conf/ccs/FredriksonJR15}. We first train the ResNet model based on the MNIST dataset with epoch = 10, batch size = 128 and learning rate = 0.1. 
Since these two evaluation schemes only support instance-level unlearning, we select the unlearning data as all instances in class three and consider other class data as the remaining data. To evaluate the performance of the retraining and fine-tuning approaches, we perform unlearning operations on the already trained model using both methods separately. For the retraining and fine-tuning unlearning process, we use the same training setting as the original training process. Subsequently, We attack the above three models (original, retrained, and fine-tuned) once per epoch based on model inversion attacks. Figure~\ref{fig:modelinversionattack} illustrates partial results of model inversion attacks against the original model and the other two models affected by different unlearning methods. 
Additionally, we use MIAs in paper~\cite{DBLP:conf/csfw/YeomGFJ18} to evaluate whether the unlearning data are still identifiable in the training dataset. We set the number of shadow models as 10 and the training epoch of the shadow model as 10, batch size = 64. The attack model is a fully connected network with two hidden linear layers of width 256 and 128, respectively, with ReLU activation functions and a sigmoid output layer. We evaluate our unlearning scheme with two settings, MNIST + ResNet and CIFAR-10 + ResNet, and set unlearning class = 3. We equally divide the training dataset into two subsets to generate the training dataset for shadow models and then train the attack model based on the output of those shadow models. Figure~\ref{fig:membershipinferenceattack} shows the results of our evaluation.

\textbf{Results.} As shown in Figure~\ref{fig:modelinversionattack}, Figure~\ref{fig:modelinversionattack_original} illustrates the results of the original training process. It shows a notable trend: as the training progresses, the model becomes increasingly knowledgeable about the unlearning class 3, leading to a more pronounced exposure of information through model inversion attacks. For the retraining scheme in Figure~\ref{fig:modelinversionattack_retraining}, the attack's effectiveness is almost ineffective since the model lacks knowledge about the unlearning class. For the fine-tuning scheme in Figure~\ref{fig:modelinversionattack_finetuning}, the results reveal that after the first round of fine-tuning, the information about the unlearning class becomes considerably challenging to recover. This is because the model inversion attack has relatively little information about the unlearning data to rely on after only one epoch fine-tuning process. As the fine-tuning process progresses, the model retains less and less information about the unlearning class. Consequently, attempts to recover those data through model inversion attacks become increasingly challenging.

For the results of MIAs in Figure~\ref{fig:membershipinferenceattack}, before the fine-tuning process~(points with x=0), MIAs have a high success rate for all original models; i.e., it successfully derived the training dataset containing unlearning class. For the retraining process, the success rate of MIAs is zero, indicating that MIAs cannot determine the existence of unlearning class in the training dataset since the unlearning class actually isn't in the training set. For the fine-tuning scheme, as the information contained in the model decreases, the effectiveness of the attack also decreases. This suggests that the fine-tuning scheme can also remove information about unlearning data from the model. In addition, we also conduct measurements on model accuracy for both the unlearning data and the remaining data during both unlearning processes with the MNIST dataset. The results are shown in Figure~\ref{fig:membershipinferenceattack_c}. During the retraining process, the model's accuracy on the unlearning data is $0$ because it didn't learn anything from this specific data. On the other hand, for the fine-tuned model, the accuracy of the unlearning data gradually decreases due to the effect of catastrophic forgetting. As for the remaining data, the fine-tuning process shows minimal changes, whereas the retraining process continually learns and improves to eventually achieve a usable state. It is important to highlight that the fine-tuning model achieves much faster unlearning results compared to retraining the model to a usable state. This suggests that the fine-tuning scheme has a better unlearning effect compared to the retraining scheme.

\textbf{Summary.} As expected, similar to the retraining scheme, the fine-tuning scheme hinders the inference of any meaningful information concerning the unlearning data. Moreover, the fine-tuning scheme proves to be more efficient compared to the retraining scheme. Therefore, in this paper, we adopt the fine-tuning scheme for the unlearning of features.

\subsection{Identifying results}
\label{section:identifying}

As discussed in Section~\ref{subsection:featureIdentificationwithoutknownannotations}, when feature annotations are unavailable, we focus on training the model to identify feature information from the image automatically. To accomplish this, we propose applying the eigengap heuristic technique, which optimizes the clustering process and enhances the extraction of additional feature information. In this Section, we provide a comparative analysis between our enhanced scheme and the original scheme in~\cite{DBLP:conf/ijcai/0002WHZFZZ21} within this section.

\begin{figure*}[!t]
      \centering
      \subfloat[Results from icCNN]{\includegraphics[width=0.5\textwidth]{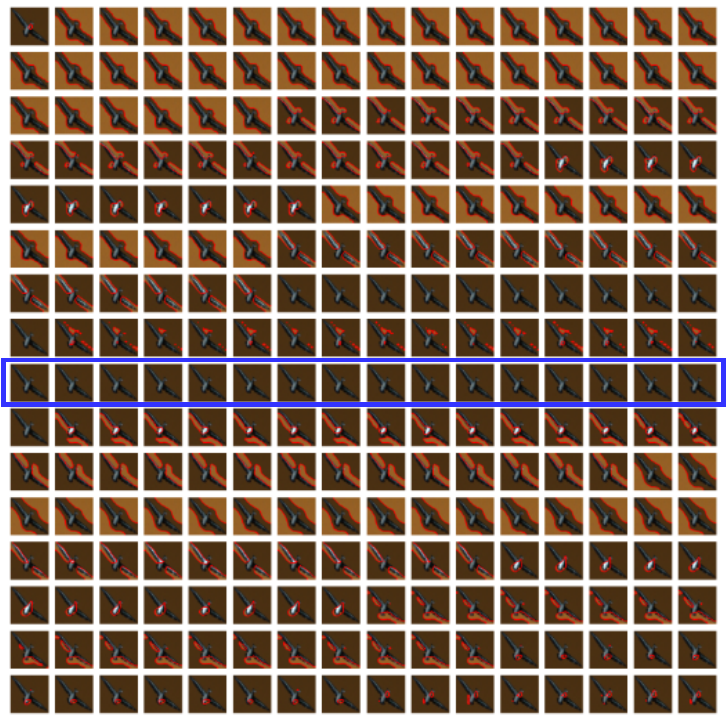}\label{fig:Featureextractionscheme_a}}
      \subfloat[Resuls from our scheme]{\includegraphics[width=0.5\textwidth] {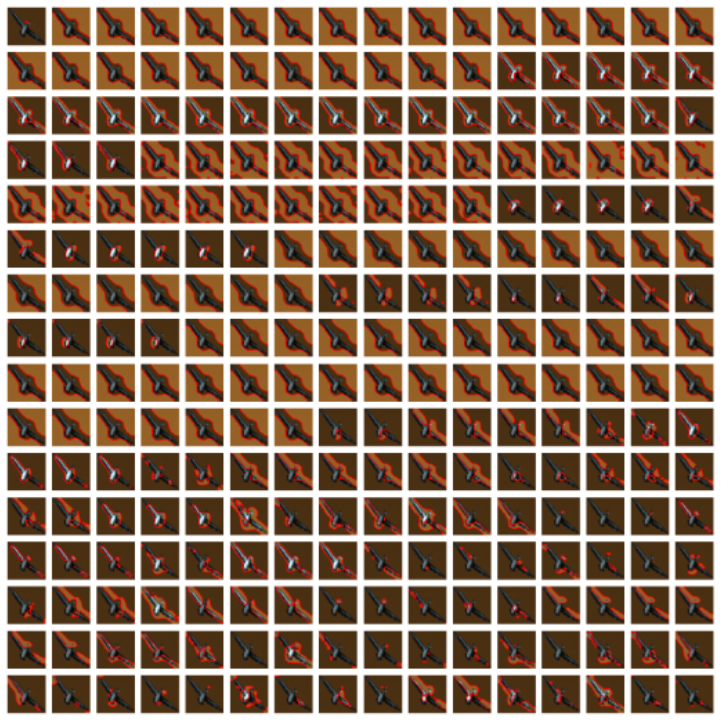}\label{fig:Featureextractionscheme_b}}
    \caption{Visualization of feature maps of icCNN~\cite{DBLP:conf/ijcai/0002WHZFZZ21} and our scheme. Both visualizations show different group filters identifying different feature patterns, such as bird wings and stomach. In the same groups, filters identify almost similar pattern features, aligning with the results in~\cite{DBLP:conf/ijcai/0002WHZFZZ21}. In addition, compared to the results in Figure~\ref{fig:Featureextractionscheme_a} from~\cite{DBLP:conf/ijcai/0002WHZFZZ21}, our method identifies more types of pattern features, and all filters are involved in the identification process, which can be used to unlearn various feature information in unlearning step.
    }%
    \label{fig:Featureextractionscheme}
\end{figure*}

\textbf{Setup.} We followed experimental settings in Shen et al.~\cite{DBLP:conf/ijcai/0002WHZFZZ21} to construct our experiment. Specifically, we train the identifier model based on ResNet architecture~\cite{DBLP:conf/cvpr/HeZRS16} and choose the bird category in the PASCAL-Part dataset~\cite{DBLP:conf/cvpr/ChenMLFUY14} to optimize our model. Just like~\cite{DBLP:conf/ijcai/0002WHZFZZ21,DBLP:journals/pami/ZhangWWZZ21}, we add the loss $L_{cls}(\mathcal{D}, \mathbf{w}, \mathbf{A})$ to the first convolutional layer after $layer3$ as our target layer. This is because filters in high convolutional layers tend to capture object parts or pattern features rather than fine-grained textures~\cite{DBLP:conf/cvpr/BauZKO017}. We first train the original ResNet model based on the classification loss $L_{ori}(\mathcal{D}, \mathbf{w})$. Then, we simultaneously optimize loss $L_{cls}(\mathcal{D}, \mathbf{w}, \mathbf{A})$ and $L_{ori}(\mathcal{D}, \mathbf{w})$ to fine-tune the trained model so that its parameters can recognize different pattern features. In the first step, we set batch size = 256, learning rate = 0.000001 and epoch = 200. In the second step, we set batch size = 16, learning rate = 0.00001 and epoch = 2500. For icCNN scheme in~\cite{DBLP:conf/ijcai/0002WHZFZZ21}, we set the cluster number as 16. For $\lambda$ in two schemes, we set 1. We followed Zhang et al.~\cite{DBLP:journals/pami/ZhangWWZZ21} to visualize each filter’s feature map in our target layer. The results are shown in Figure~\ref{fig:Featureextractionscheme}.

\textbf{Results.} In Figure~\ref{fig:Featureextractionscheme}, Figure~\ref{fig:Featureextractionscheme_a} shows the identification results of the target layer from the icCNN model~\cite{DBLP:conf/ijcai/0002WHZFZZ21}, while Figure~\ref{fig:Featureextractionscheme_b} shows the results from our identifier model. Both visualization results indicate that each filter in different groups can identify various feature information, such as the wings and stomach of the bird, while in the same groups, each filter identifies almost the same feature. This result is the same as that in~\cite{DBLP:conf/ijcai/0002WHZFZZ21}. In addition to this, compared to Figure~\ref{fig:Featureextractionscheme_a} from~\cite{DBLP:conf/ijcai/0002WHZFZZ21}, our method recognizes more types of feature information and all the filters are involved in the identification operation. For~\cite{DBLP:conf/ijcai/0002WHZFZZ21}, on the other hand, since the number of clusters is randomly determined and does not consider whether the resulting clustering is an optimal solution, it leads to some filters not participating in the identification process~(Filters as marked in the blue box in Figure~\ref{fig:Featureextractionscheme_a}). This comparison illustrates that our scheme can achieve better optimization results and identify more pattern features,  which can be used to unlearn various feature information in the unlearning step.

\subsection{The effect of hyper-parameters}
\label{sec:effectparameters}
\begin{figure}[!t]
      \centering
      \subfloat[Evaluation the effects of $\lambda$]{\includegraphics[width=0.45\textwidth]{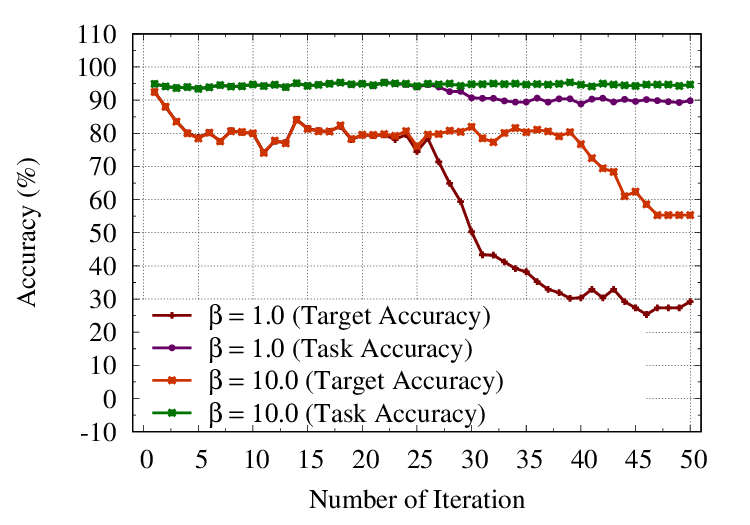}\label{fig:hypyerparameters_a}}\\
      \subfloat[Evaluation the effects of $\beta$]{\includegraphics[width=0.45\textwidth]{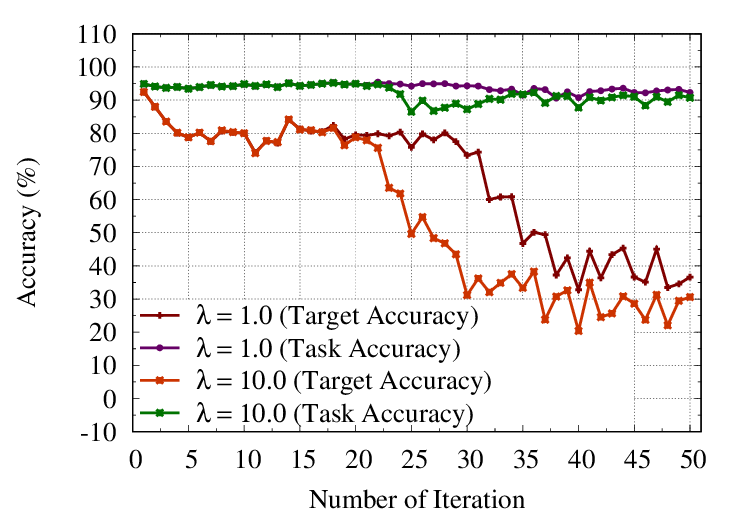}\label{fig:hypyerparameters_b}}
    \caption{The Effect of Hyper-Parameters. In Figure~\ref{fig:hypyerparameters_a}, when fixing $\lambda$, a smaller $\beta$ unlearns the target feature quickly but impacts the original model's accuracy. In Figure~\ref{fig:hypyerparameters_b}, a larger $\lambda$ speeds up feature unlearning but reduces the original model's performance.}
    \label{fig:hypyerparameters}
\end{figure}

\textbf{Setup.} In our feature unlearning with known annotations, there are two hyperparameters: $\lambda$ and $\beta$. To evaluate the impact of these hyperparameters, we set the following experiments: We opt for the ResNet architecture for both the original and adversary models, while we select the U-Net architecture for the remover model. In the initial step, we separately train the original model and the adversary model to identify the tasks of `Smiling' and `Mouth Slightly Open' using the CelebA dataset. Our goal is to unlearn information related to whether the mouth is open, specifically the `Mouth Slightly Open' feature, from the original model. During the training process, we set the following parameters: epoch = 10, batch size = 50, and learning rate = 0.000005. For the adversarial unlearning process, we set the batch size to 36 and the learning rate = 0.000005. We hold one hyperparameter constant while allowing the other hyperparameter equal to  $1.0$ and $10.0$, respectively. We then record the model accuracy after a single iteration of adversarial unlearning. Figure~\ref{fig:hypyerparameters} shows the results.

\textbf{Results.} It can be seen from Figure~\ref{fig:hypyerparameters_a}, when fixing $\lambda$, a smaller $\beta$ unlearns the target feature quickly but affects the accuracy of the original model. In Figure~\ref{fig:hypyerparameters_b}, a larger $\lambda$ speeds up the feature unlearning of the model, but again reduces the performance of the original model. In summary, larger $\lambda$ and smaller $\beta$ will affect the model performance for the original model. In practice, it should choose appropriate values to achieve unlearning while minimizing the impact on the performance of the original model.

\subsection{Feature unlearning application}
\label{section:featureunlearningapplication}
As detailed in Section~\ref{sec:introduce}, the concept of feature unlearning can be regarded as an alternative approach to address fairness issues within models. In this Section, we proceed to evaluate the efficacy of implementing feature unlearning as a strategy to alleviate biases inherent in models.

\begin{figure*}[!t]
      \centering
      \subfloat[Unlearning result for Mouth (Male) ]{\includegraphics[width=0.33\textwidth]{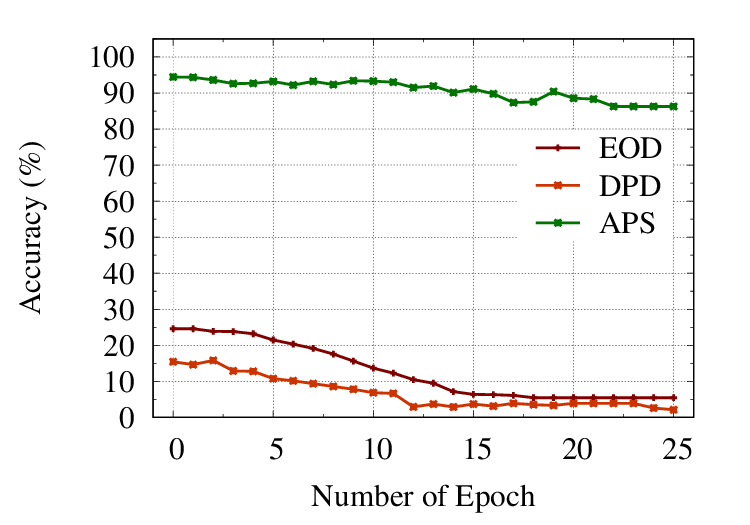}\label{fig:fairnessevaluation_a}}\hfill
      \subfloat[Unlearning result for Young (Smiling)  ]{\includegraphics[width=0.33\textwidth]{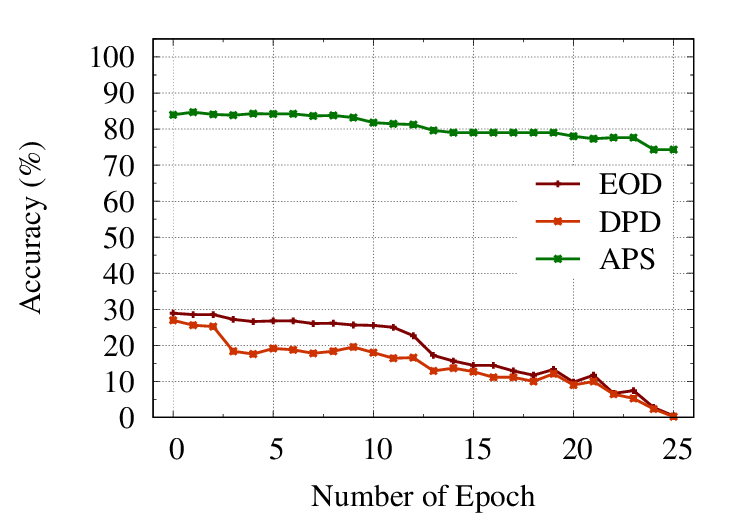}\label{fig:fairnessevaluation_b}}\hfill
      \subfloat[Unlearning result for Bald (Mouth)  ]{\includegraphics[width=0.33\textwidth]{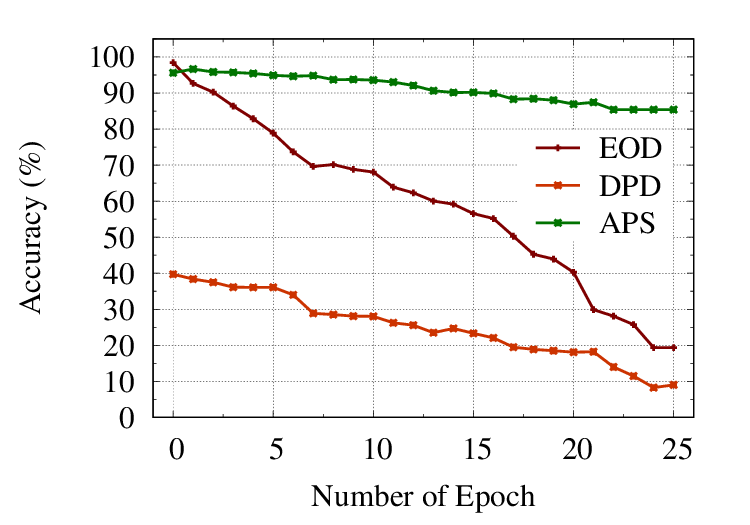}\label{fig:fairnessevaluation_c}}\hfill \\
      \subfloat[Naive result for Mouth (Male) ]{\includegraphics[width=0.33\textwidth]{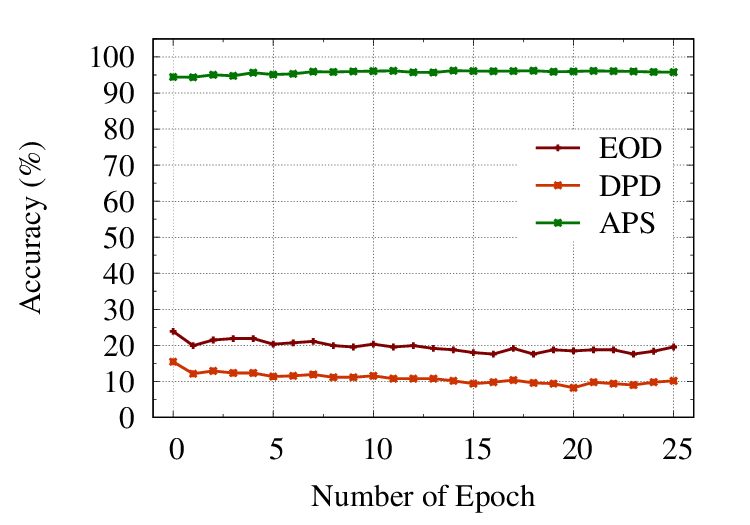}\label{fig:fairnessevaluation_d}}\hfill
      \subfloat[Naive result for Young (Smiling)  ]{\includegraphics[width=0.33\textwidth]{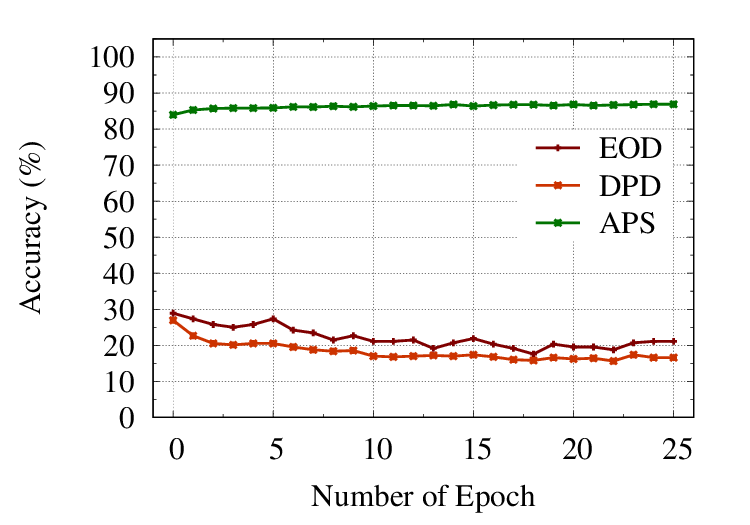}\label{fig:fairnessevaluation_e}}\hfill
      \subfloat[Naive result for Bald (Mouth)  ]{\includegraphics[width=0.33\textwidth]{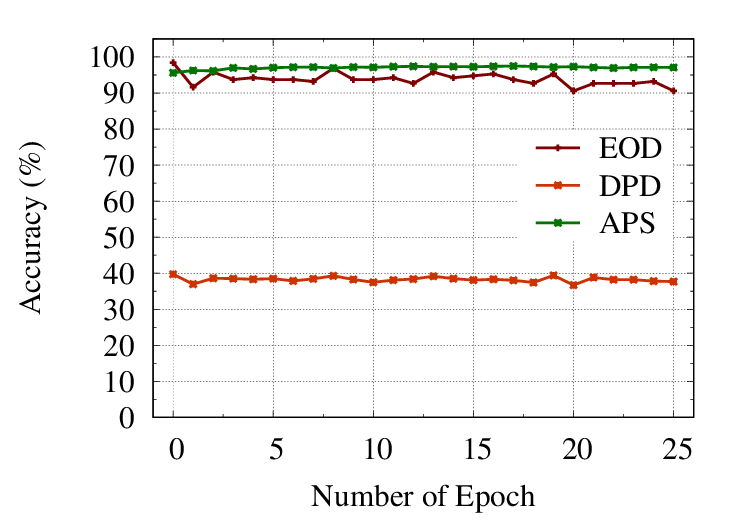}\label{fig:fairnessevaluation_f}}\hfill \\
    \caption{The results of debiasing based on feature unlearning. During the unlearning process, values for EOD and DPD gradually decrease, indicating bias reduction in the model. All naive approaches didn't reduce bias. This suggests feature unlearning retains task-related features while removing bias-associated features. All results align with our hypothesis that feature unlearning can be used as an optional strategy to alleviate biases inherent in models.
    }
    \label{fig:fairnessevaluation}
\end{figure*}

\textbf{Setup.} We consider three distinct debiasing scenarios that cover different combinations of pattern features and attribute features: the removal of gender bias (specifically, male bias) from a model classifying whether a mouth is slightly open; secondly, the alleviating of smiling bias from a model categorizing whether an individual is young and lastly, the mitigation of bias associated with mouth slightly open in a bald classification model. We reconstruct the dataset to simulate the bias present in the dataset. The configuration of the reconstructed dataset is shown in Table~\ref{tab:fairnessconfiguration}. We set epoch=10, batch size =50, and the learning rate=0.000005 for training the adversary model. For the original model, we set epoch = 50, batch size = 128 and learning rate = 0.001. The setting for each feature unlearning process is illustrated in Table~\ref{tab:fairnesssetting}. To better illustrate that the bias in the model is indeed alleviated by the unlearning process, we set up a comparison experiment, i.e., we only use the same data to fine-tune the model without the unlearning process~(denoted as naive method). If the unlearning scheme succeeds in alleviating bias and the naive method does not alleviate bias, this suggests that feature unlearning can be used as an optional program for removing bias from the model. We use the \textit{equalized odds difference}~(EOD) and \textit{demographic parity difference}~(DPD) in the \textit{fairlearn.metrics} package to evaluate the model bias and use the \textit{average precision score}~(APS) to test the model performance. The results are shown in Figure~\ref{fig:fairnessevaluation}.

\textbf{Results.} As shown in Figure~\ref{fig:fairnessevaluation}, Figure~\ref{fig:fairnessevaluation_a} to Figure~\ref{fig:fairnessevaluation_c} illustrate the results of alleviating bias based on feature unlearning, while Figure~\ref{fig:fairnessevaluation_d} to Figure~\ref{fig:fairnessevaluation_f} show the results from the naive scheme. The values corresponding to \textit{equalized odds difference}~(EOD) and \textit{demographic parity difference}~(DPD) exhibit a gradual diminution during the unlearning process, signifying progressive alleviation of bias within the model. Conversely, the naive scheme did not reduce bias within the model. Furthermore, the strategy of feature unlearning demonstrates a minimal impact on the model's accuracy, as evidenced by the marginal alteration in the \textit{average precision score}~(APS) value. This observation implies that the process of feature unlearning effectively preserves feature information associated with the model task and removes all information about the feature that creates bias. This observation aligns seamlessly with our initial hypothesis that feature unlearning is an elective approach for alleviating model bias.

\begin{table}[!t]
\caption{The configuration of training biased model.}
\label{tab:fairnessconfiguration}
\centering
    \renewcommand{\arraystretch}{1.3}
    \begin{tabular}{cccc}
    \hline
    \rowcolor{gray!20}                  & Mouth (Male)   & Young (Smiling)  & Bald (Mouth)\\ \hline
    Both False                          & 2000           & 2000             & 2000    \\
    \rowcolor{gray!20}False and True    & 16000          & 10000            & 10000     \\
    True and False                      & 16000          & 10000            & 10000     \\
    \rowcolor{gray!20}Both True         & 2000           & 2000             & 2000     \\ \hline
    \end{tabular}
\end{table}

\begin{table}[!t]
\caption{The experimental setting for debiasing.}
\label{tab:fairnesssetting}
\centering
    \renewcommand{\arraystretch}{1.3}
    \begin{tabular}{cccc}
    \hline
    \rowcolor{gray!20}                      & Mouth (Male)   & Young (Smiling)    & Bald (Mouth)\\ \hline
    Epoch                                   & 25             & 25                 & 25    \\
    \rowcolor{gray!20}Batch Size            & 64             & 64                 & 64     \\
    LR (remover)                         & 0.00001        & 0.00001            & 0.000025     \\
    \rowcolor{gray!20}LR (fine-tuning)    & 0.00001        & 0.000005           & 0.00001     \\
    $\lambda$                               & 1.6            & 1.3                & 1.6     \\
    \rowcolor{gray!20}$\beta$               & 1.0            & 1.0                & 1.0     \\\hline
    \end{tabular}
\end{table}

\section{Conclusion and Future Work}
\label{sec:conclusion}
In this paper, we have proposed an innovative machine unlearning approach that enables the selective removal of feature information from a trained model. We consider two types of unlearning requests: feature
unlearning with known annotations and feature unlearning without annotations. In the case of unlearning with known annotations, we utilize adversarial learning to eliminate feature-related information from the model. For unlearning without annotations, we design a re-encoded and fine-tuning technique. The experimental results provide evidence that our approach effectively enables the model to eliminate the impact of features while maintaining accuracy in a quick and efficient manner.

Future work will focus on extending or modifying current methods to make it applicable to other complex types of features, such as the texture of the object or the sentiment of the language. In addition, evaluation schemes for feature unlearning also need further research. Furthermore, we have future plans to develop a more comprehensive scheme capable of effectively addressing unlearning requests from Natural Language Processing (NLP) or Generative Adversarial Networks (GANs). 

\bibliographystyle{IEEEtran}
\bibliography{sampleBibFile}

\end{document}